\documentclass[iop]{emulateapj}
\usepackage{color}
\usepackage{lscape}

\newcommand{\HI}{\ion{H}{1}~}
\newcommand{\HIstop}{\ion{H}{1}.}

\newcommand{\kms}{km s$^{-1}$~}

\slugcomment{Accepted for publication in ApJ on Aug 25, 2015}

\shorttitle{Faint Atomic Gas in HCGs}
\shortauthors{Borthakur et al.}

\begin{document}

\title{Distribution of Faint Atomic Gas in Hickson Compact Groups}

\author{Sanchayeeta Borthakur}
\affil{Department of Physics and Astronomy, Johns Hopkins University, Baltimore, MD 21218, USA}
\email{sanch@pha.jhu.edu}

\author{Min Su Yun}
\affil{Astronomy Department, University of Massachusetts, Amherst, MA 01003, USA}

\author{Lourdes Verdes-Montenegro}
\affil{Instituto de Astrof\'{\i}sica de Andaluc\'{\i}a, CSIC, Apdo. Correos 3004, E-18080 Granada, Spain}

\author{Timothy M. Heckman, Guangtun Zhu}
\affil{Department of Physics and Astronomy, Johns Hopkins University, Baltimore, MD 21218, USA}

\author{James A. Braatz}
\affil{National Radio Astronomy Observatory, 520 Edgemont Road, Charlottesville, VA 22903, USA}

\begin{abstract}

We present 21cm \HI observations of four Hickson Compact Groups with evidence for a substantial intragroup medium using the Robert C. Byrd Green Bank Telescope (GBT)\footnote{The National Radio Astronomy Observatory is a facility of the National Science Foundation operated under cooperative agreement by Associated Universities, Inc.}. By mapping H I emission in a region of 25$^{\prime}\times$25$^{\prime}$ (140-650~kpc) surrounding each HCG, these observations provide better estimates of \HI masses. In particular, we detected 65\% more \HI than that detected in the Karl G. Jansky Very Large Array (VLA) imaging of HCG~92. We also identify if the diffuse gas has the same spatial distribution as the high-surface brightness (HSB) H I features detected in the VLA maps of these groups by comparing the \HI strengths between the observed and modeled masses based on VLA maps. We found that the H I observed with the GBT to have a similar spatial distribution as the HSB structures in HCGs~31 and 68. Conversely, the observed \HI distributions in HCGs~44 and 92 were extended and showed significant offsets from the modeled masses. Most of the faint gas in HCG~44 lies to the Northeast-Southwest region and in HCG~92 lies in the Northwest region of their respective groups. The spatial and dynamical similarities between the total (faint+HSB) and the HSB HI indicate that the faint gas is of tidal origin. We found that the gas will survive ionization by the cosmic UV background and the escaping ionizing photons from the star forming regions and stay primarily neutral for at least 500~Myrs.

\end{abstract}

\section{INTRODUCTION \label{Sec:intro}}

Hickson Compact Groups (HCGs) are dense concentration of 4-10 galaxies that have densities similar to galaxy clusters, but are relatively isolated \citep{hickson82}. Their low velocity dispersions ($\Delta V \sim$200~\kms; see \citet{hickson92}) and very high galaxy densities favor extreme tidal interactions. Such interactions alter the morphology and distribution of atomic gas in their host galaxies and have been studied by extensively \citep{willvan95, VM01, williams02,VM02, VM05, durbala08,demello12}. 
The atomic gas in some of these groups exists in complex tidal structures and even spatial offsets from their host galaxies. The groups are also found to be deficient in \HI  \footnote{This is referred to as ``\HI-deficiency'' and is defined as $Def_{HI}\equiv$ log[M(HI)$_{pred}$] $-$ log[M(HI)$_{obs}$] \citep[see ][ for predicted HI masses]{haynes84}.} \citep{williams87, huchtmeier97}.

\begin{figure*}
 \figurenum{1}
 \hspace{-0.5cm}
\includegraphics[trim=12mm 14mm 16mm 26mm, clip=true, scale=0.594,angle=-0]{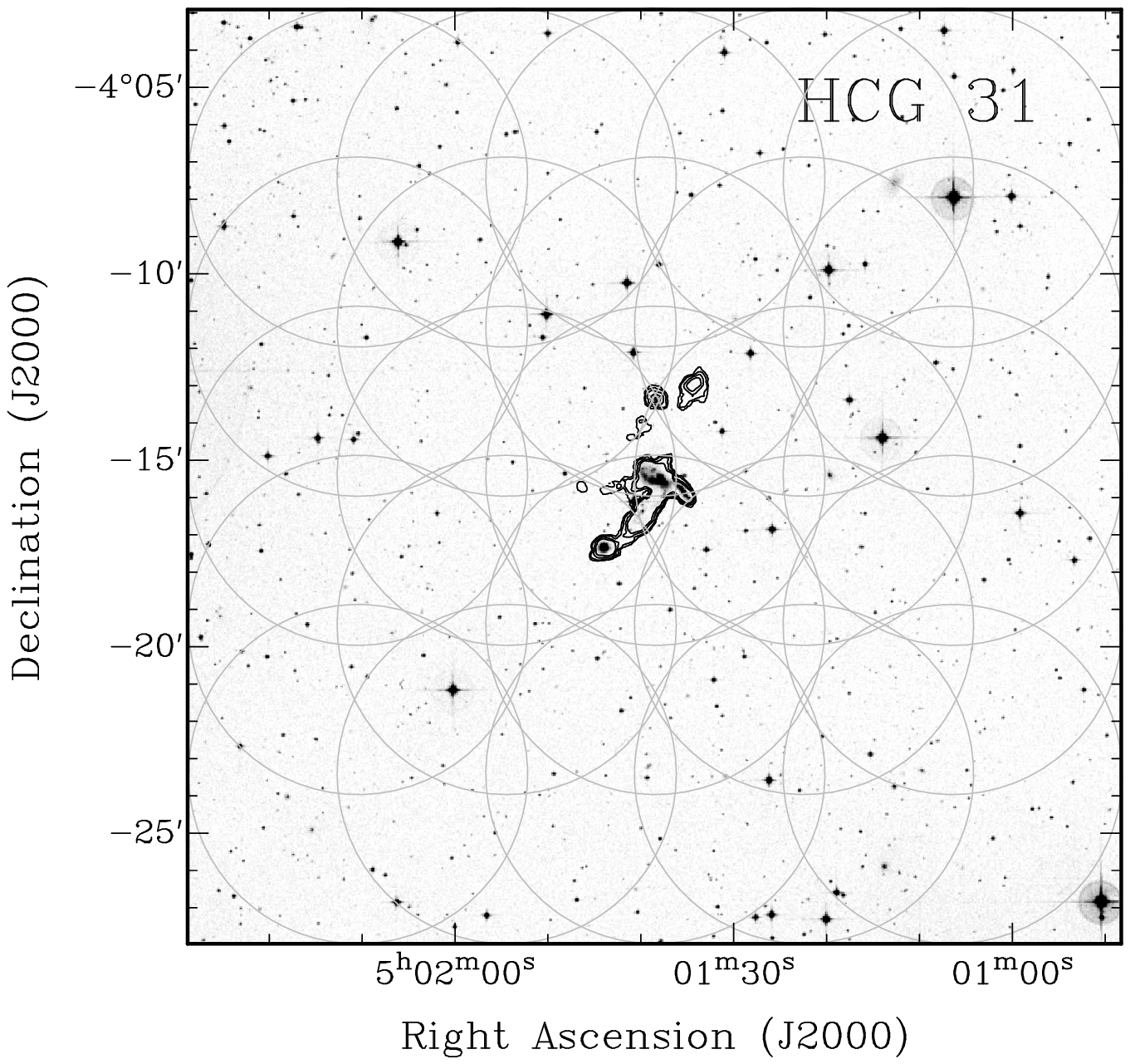} 
\includegraphics[trim=12mm 10mm 16mm 24mm, clip=true, scale=0.57,angle=-0]{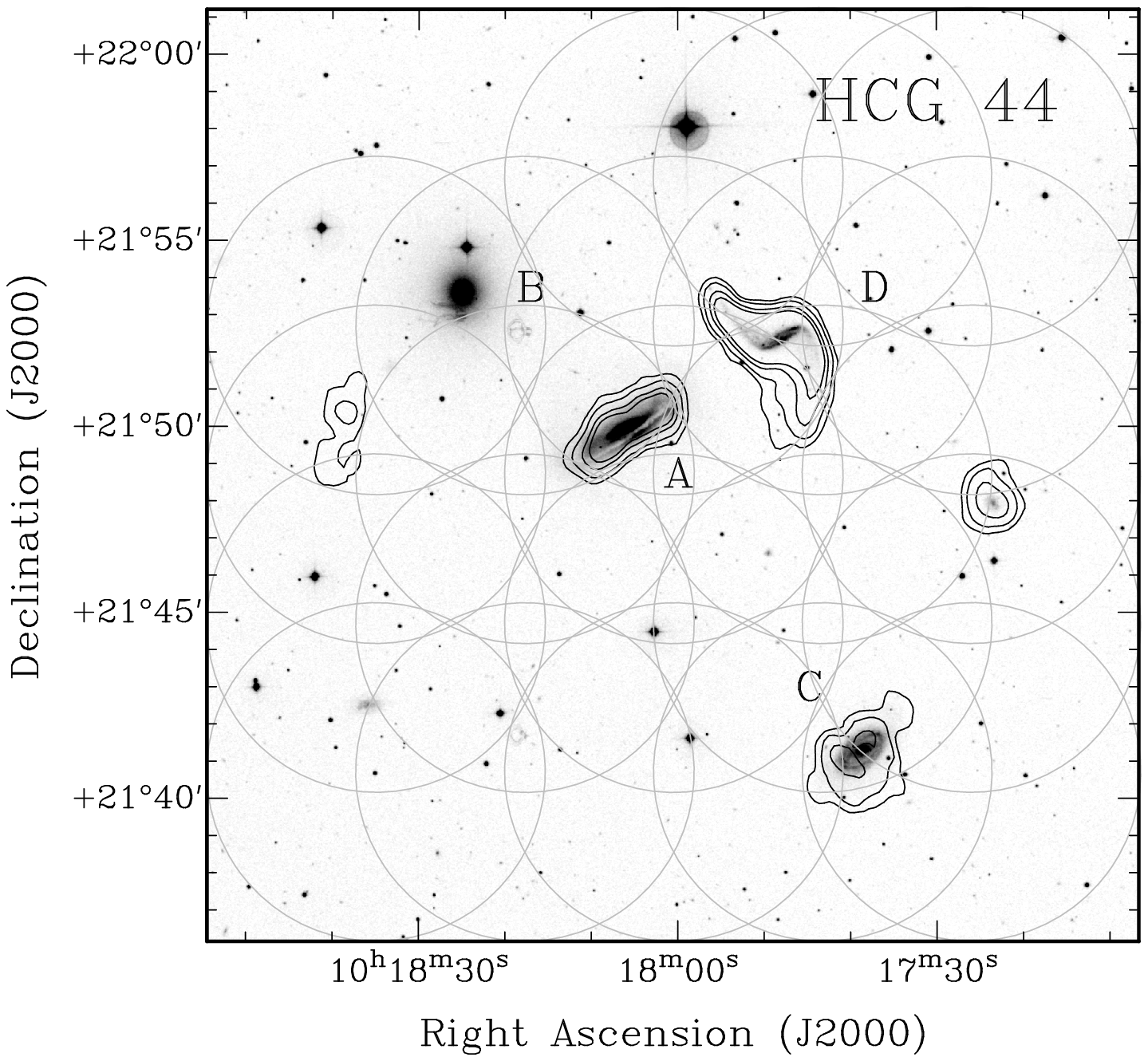} \\

 \hspace{-0.5cm}
\includegraphics[trim=12mm 15mm 16mm 30mm, clip=true, scale=0.57,angle=-0]{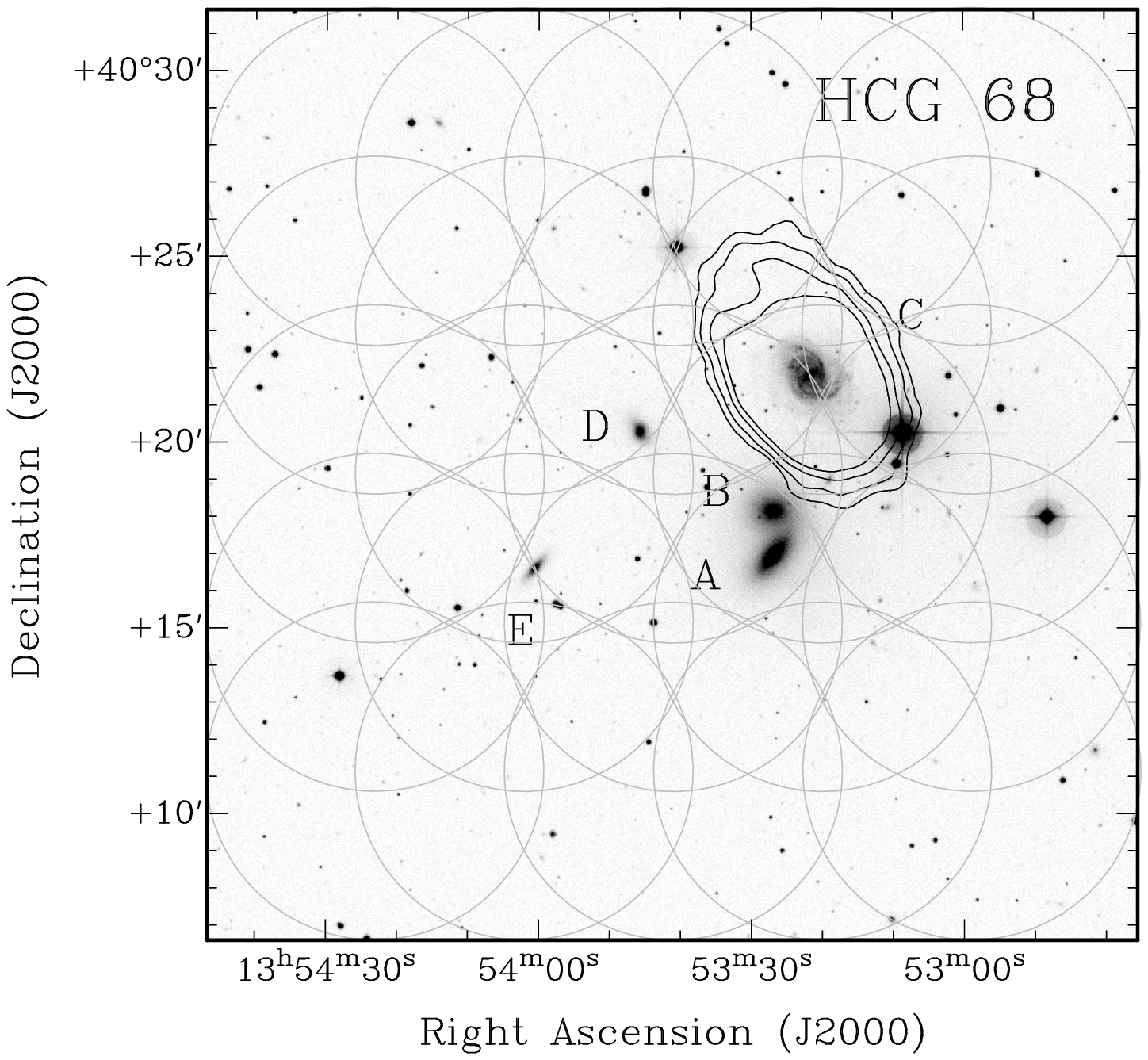} 
\includegraphics[trim=12mm 70mm 16mm 60mm, clip=true, scale=0.58,angle=-0]{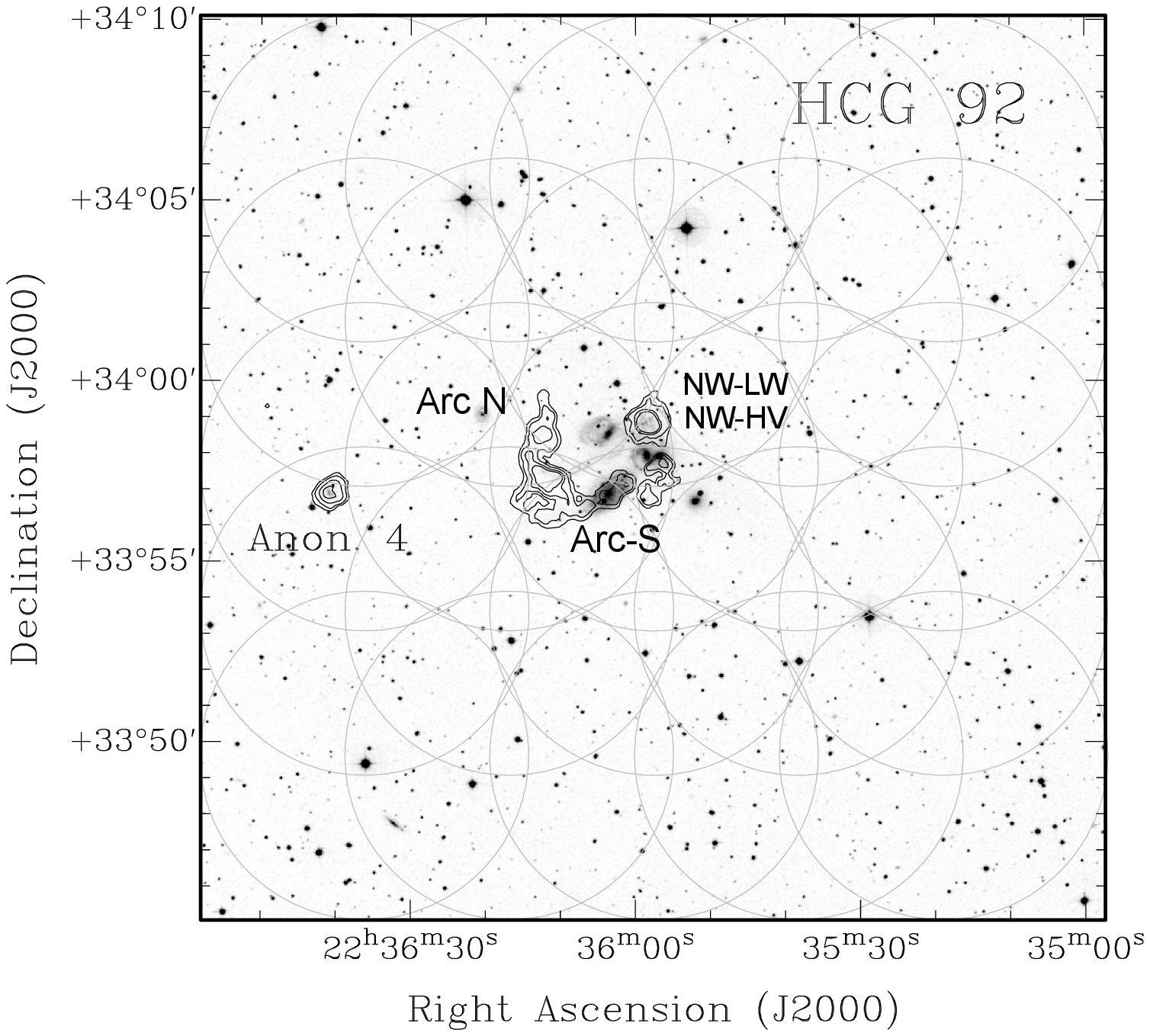} \\
\caption{Digital Sky Survey r-band images of HCGs~31, 44, 68, and 92 in greyscale with 21~cm VLA \textsc{Hi} maps overlaid as black contours. The member galaxies as defined by \citet{hickson82} are labeled for extended groups (HCG~44 and 68). The  GBT beam size (FWHM=9.1$^\prime$) for our pointings are overplotted in grey.
\label{fig-vla_maps}}
\end{figure*}

\citet[][VM01 hereafter]{VM01} suggested that the nature and distribution of \HI in the tidal structures can be used as a proxy to estimate the level of interaction and hence the `phase of evolution' of the group. They proposed that the groups in their early stages of evolution would have galaxies with individual \HI\ envelopes. As the group evolves, the tidal interactions would increase and result in severe tidal stripping. As a result, most of the \HI\ in such systems should exist as tidal debris. 
In the final stages of evolution the stripped \HI\ gas from multiple galaxies gets stirred-in together. This blends the individual clumps and facilitates the formation of large \HI envelopes around the groups. It is also likely that during the process of interactions part of the neutral gas goes through a phase change via ionization, thus making the evolved groups deficient in \HIstop

 \begin{figure*}
 \figurenum{2}
 \hspace{-0.5cm}
\includegraphics[scale=0.56,angle=-0]{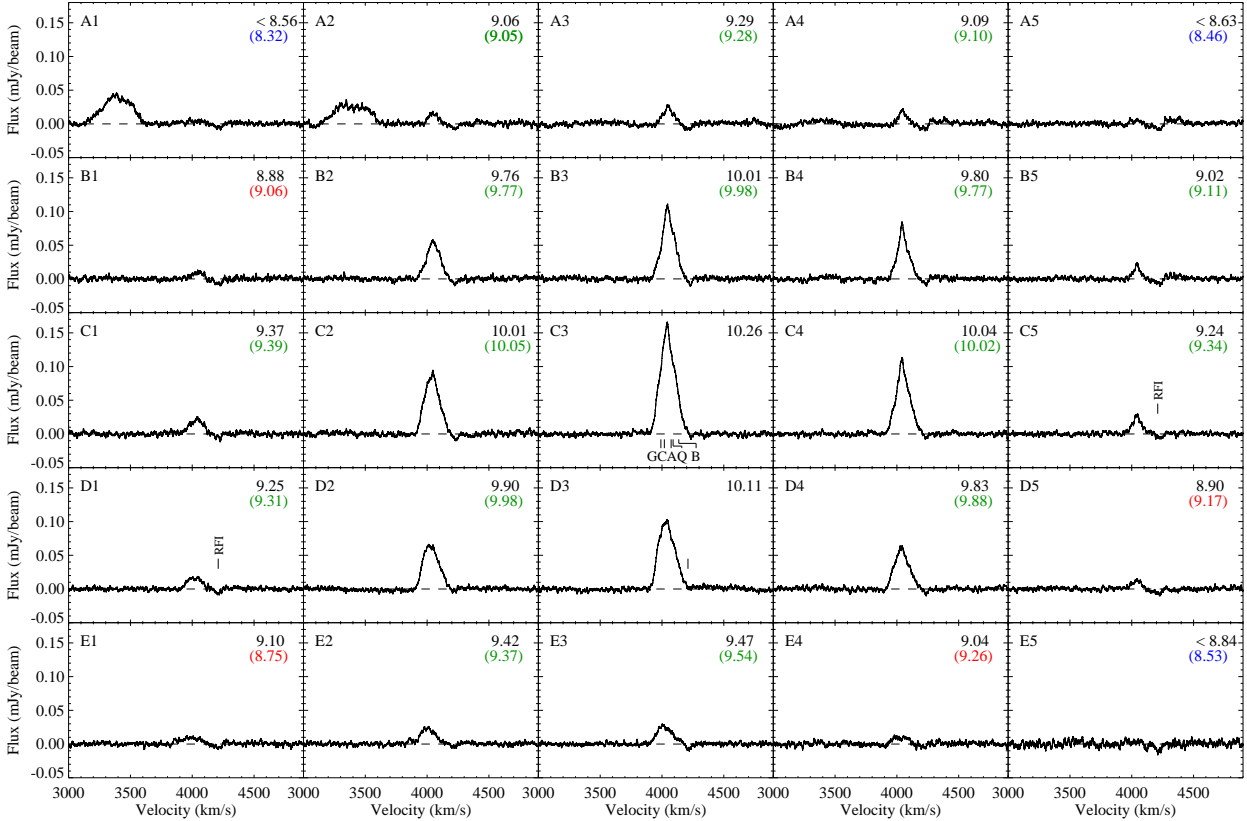} 
\caption{GBT 21cm HI spectra for each of the four HCGs in our sample. The spectra have been plotted in a grid format similar to the 5x5 pointing scheme on the sky. Pointings are separated by 4$^\prime$ from their neighbors resulting in a total coverage of 25$^\prime$x25$^\prime$.  The center of the map coincides with the center of the group. The data presented in this paper have been smoothed to 10~\kms. The position of the pointings are marked at the top left corner and their observed HI masses at the top right corner. We have listed the 5$\sigma$ HI masses for spectra where no HI was detected. The modeled HI masses are shown in color. They are color-coded to show the agreement/disagreement between the observed and modeled masses in green/red. Pointings where the limiting \HI masses are consistent with the predicted masses are shown in blue.  The systemic velocity of the member galaxies are marked.  {\bf (a)} 25 spectra towards HCG~31. The depression centered at 4200~\kms is an artifact left over after clearing RFI at that frequency. The group size as defined by \citet{hickson82} and  re-defined by \citet{VM05} places the entire group well within the central pointing. The redshifts of the member galaxies place all of them within the velocity coverage of the HI profile. At the redshift of the group, each pointing is separated by 66~kpc. The broad feature between 3170-3600~\kms has no associated optical galaxy. It may be a RFI feature although not clear from the current observations. Further observations are required to confirm it.  \label{fig-data}}
\end{figure*}

In a study by \citet[][B10 hereafter]{borthakur10a}, the authors compared \HI data obtained with the Karl G. Jansky Very Large Array (VLA) to that from the Robert C. Byrd  Green Bank Telescope (GBT). They found that the groups have a significant fraction of \HI  in a faint and diffuse phase that was missed by the VLA. They also found that the fraction of \HI in the diffuse component increases with the evolutionary phase of the group. 
They advocated that with increasing interaction time as marked by the evolutionary phase of the group, the tidally-stripped \HI structures disperses more and more into a faint \HI medium, which finally may even get ionized. This was identified as a contributing factor for the observed \HI deficiency.

 \begin{figure*}
 \figurenum{2}
 \hspace{-0.5cm}
\includegraphics[scale=0.56,angle=-0]{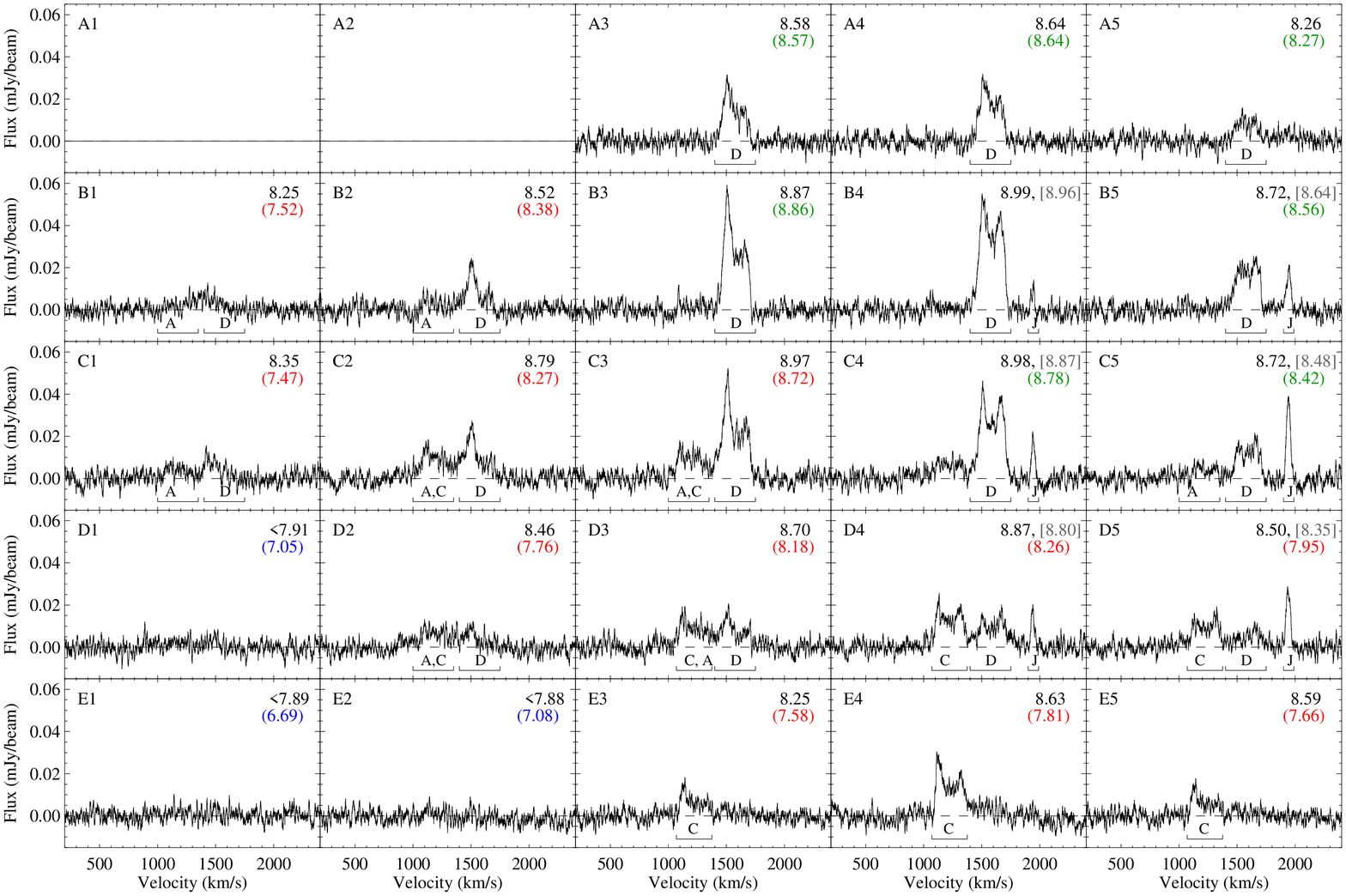} 
\caption{(b) Spectra from 23 pointings around HCG~44. Pointings are separated by $4^\prime$ or 23~kpc at the redshift of the group. We were unable to obtain spectra for two of the pointings due to time constraints during observations. Features associated with the individual galaxies are identified and are marked A, C, D, or J corresponding to 44A, 44C, 44D, and SDSS J1012. }
\end{figure*}

\begin{figure*}
 \figurenum{2}
 \hspace{-0.5cm}
\includegraphics[scale=0.56,angle=-0]{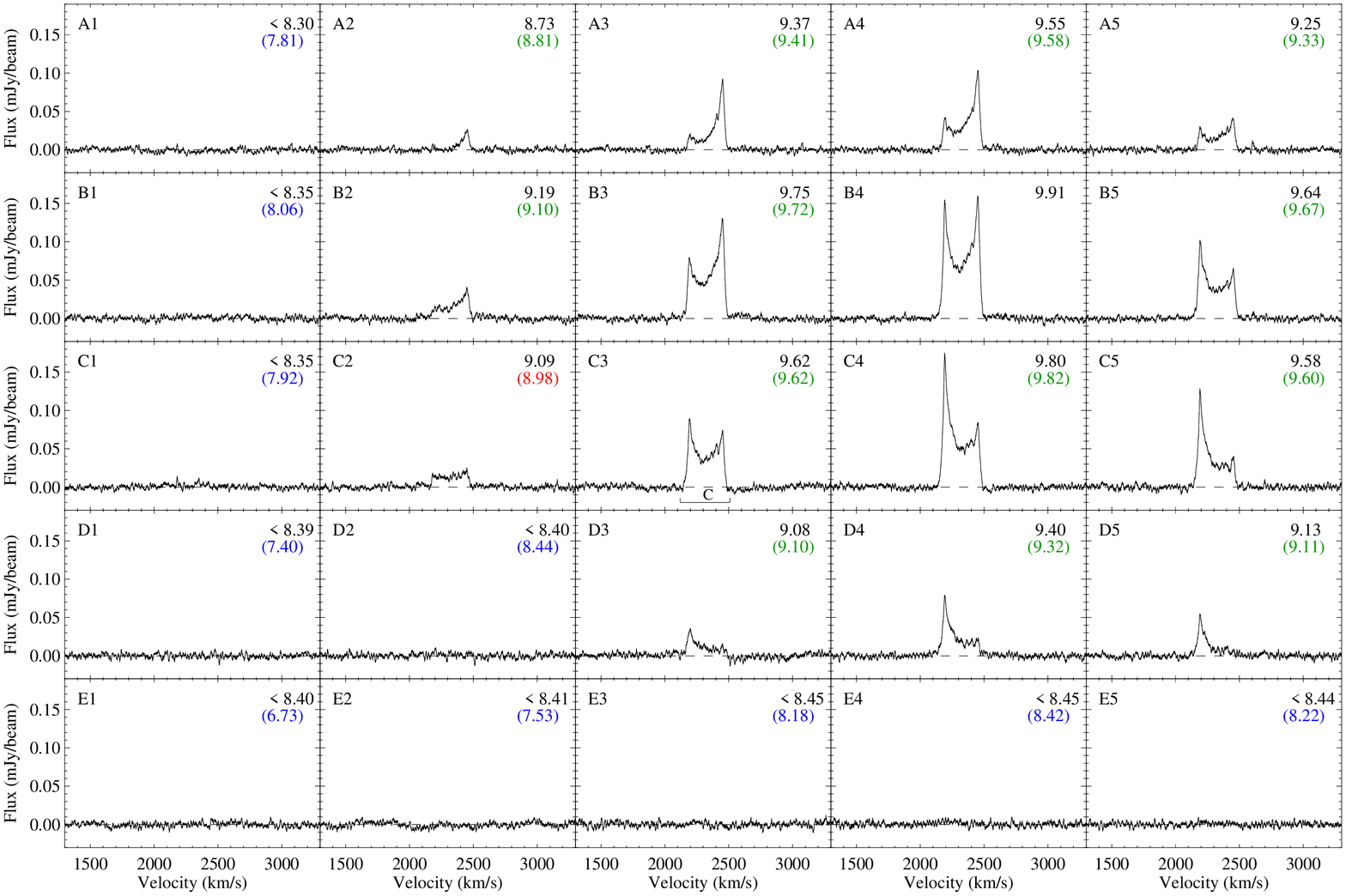} 
\caption{(c) Spectra showing the \textsc{Hi} distribution around HCG~68. The pointings are separated by 4$^{\prime}$ or 38~kpc at the redshift of the group. The only galaxy with HI in this group is HCG~68C. It shows an unperturbed doubled-horned profile and has a large extended disk of $\sim$10$^\prime$ in the VLA image. The peak of the profile centers at position B4, which is closest pointing to the position of 68C.  } 
\end{figure*}

\begin{figure*}
 \figurenum{2}
 \hspace{-0.5cm}
\includegraphics[scale=0.56,angle=-0]{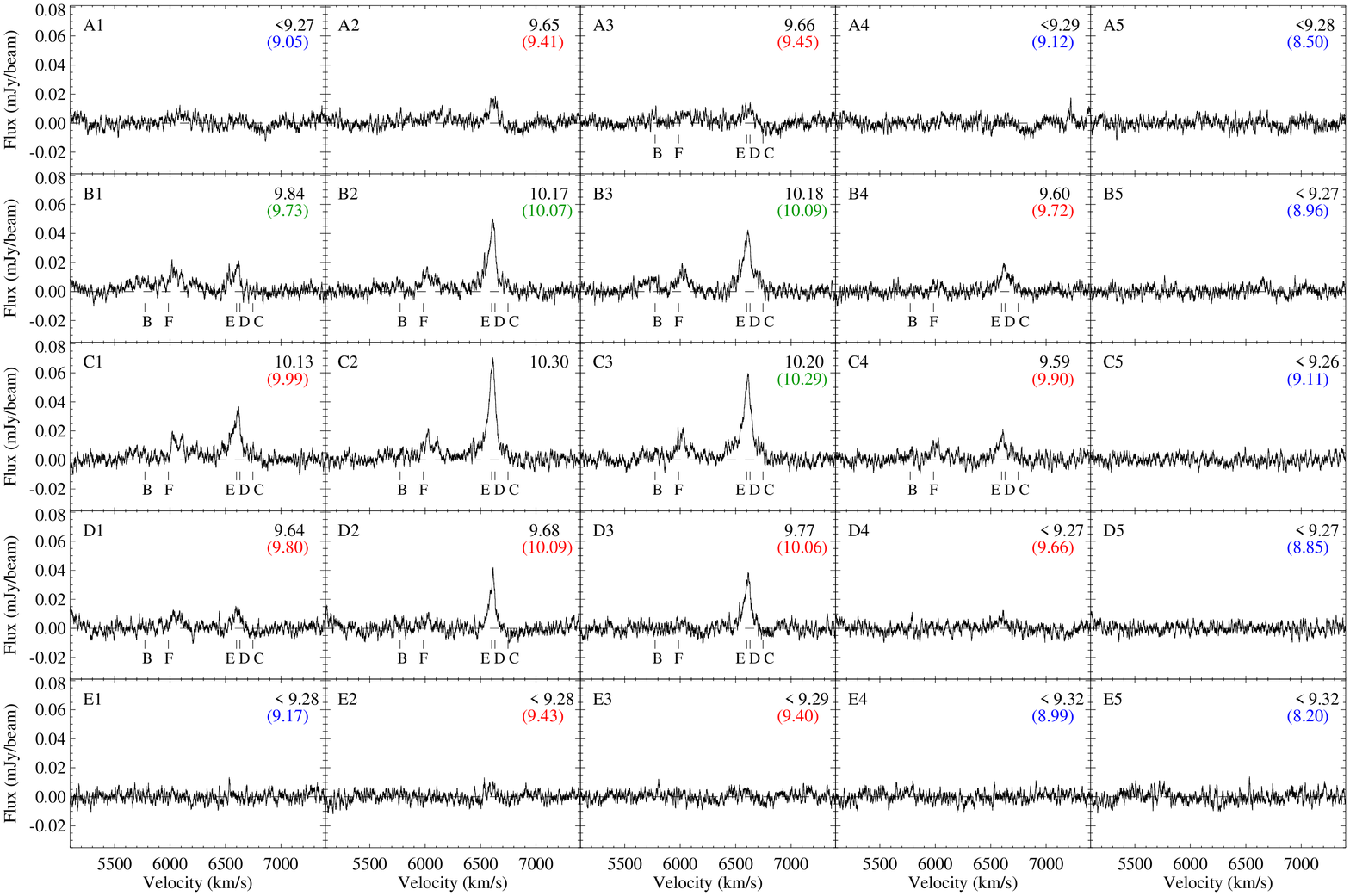} 
\caption{(d) Spectra showing the \textsc{Hi} distribution around HCG~92. The pointings are separated by 4$^{\prime}$ or 107~kpc at the redshift of the group. The redshift of the member galaxies are marked The peak flux was observed at position C2 and the observed \textsc{Hi} mass at C2 was used to normalize the modeled masses.  }
\end{figure*}

The presence of substantial amount of neutral gas in the intragroup medium can have long-lasting influence on the evolution of the group. The high neutral column densities and pressures resulting from tidal dynamics have been suggested as a potential cause for inducing star formation in tidal debris \citep{mullan13}.
Several studies have found large numbers of young stellar clusters being formed in the intragroup medium \citep[][and citations therein]{hunsberger96,  torres-flores09, gallagher10}.  For example the intergalactic star-forming regions in Stephan's Quintet appear to be devoid of older ($\rm > 10^9$~yrs) stellar populations and their SED resembles that of dusty star-forming regions in the galactic disks than dwarf star-forming galaxies \citep{Boquien10}. Although, the diffuse gas detected by B10 is unlikely to be the sites of star formation, nevertheless the diffuse gas may eventually feed star forming regions in the group. On the other hand, the starburst-driven winds can significantly impact the morphology of the optical and the \HI structures and may even produce extreme differences between the \HI and optical tidal morphologies \citep{hibbard00}. The winds will be most effective against the low column density diffuse gas and can ionize it thus enhancing the \HI deficiency. Therefore, it is crucial to understand the distribution and conditions in the diffuse gas to predict its fate and its role in the overall evolution of the group.

Motivated by these studies, we present our pilot study of the extent of the \HI\ in four HCGs using the GBT.
Based on results by B10, we infer that all the four groups have a significant fraction of their \HI\ in the faint diffuse extended phase thus making them ideal candidates for studying the extended nature of \HIstop ~We start by presenting our sample followed by details of the observations, and data analysis in Section 2. The results and a discussion of their implications are presented in Section 3. Finally, we conclude our findings in section 4 and comment on the scope of a future study. The cosmological parameters used in this study are $H_0 =70~{\rm km~s}^{-1}~{\rm Mpc}^{-1}$, $\Omega_m = 0.3$, and $\Omega_{\Lambda} = 0.7$.

\section{OBSERVATION \label{Sec:uvobservations}}

\subsection{Sample \label{Sec:sample}}

Our sample for the pilot study consists of four groups: HCGs~31, 44, 68, and 92\footnote{HCG~92 is known as Stephan's Quintet. Our observations do not include the foreground galaxy HCG~92A (aslo known as NGC 7320) at a redshift of 0.002622 ($v=\rm786~$\kms)}. 
These four groups were part of the original sample of B10 and were found to have a substantial fraction of their \HI\ in the faint diffuse extended phase. All of the groups have at least 4 spectroscopically confirmed members and are within 100~Mpc of the Milky Way. An advantage of this sample is that these groups have extensive \HI\ data including single dish observations by \citet{williams87}, \citet{huchtmeier97}, and B10 as well as 21cm \HI imaging with the VLA by \citet{williams91, williams02, VM05,tang08} and Westerbork Synthesis Radio Telescope (WSRT) by \citet{serra13}.
The properties of the sample including their redshifts, distances, group sizes, our GBT map centers, physical scales corresponding to 1$^\prime$, and references to published VLA \HI images are presented in Table~1. 
The optical images showing VLA \HI\ contours are presented in Figure~\ref{fig-vla_maps}.

The \HI distribution of these groups can be classified into two distinct morphologies.
The \HI\ in HCGs~31 and 92 exists as complex tidal structures indicative of intense tidal interactions in these groups. VLA imaging showed that a large fraction of the \HI\ in these two groups is not associated with individual galaxies (sometimes even spatially offset from the galaxies), but exists as tidal debris in the group environment. Therefore, hereafter we refer to them as tidal debris-dominated groups.
 The \HI in HCG~44 and 68 is mostly concentrated as interstellar medium (ISM) of individual member galaxies and hence they will be referred to as ISM-dominated groups for the remainder of the paper. It is worth noting that even the ISM-dominated groups are deficient in \HI, although the \HI\ in the VLA images does not show any major tidal structures. This is probably indicative of a quick phase change of the neutral material in the  intragroup medium (IGrM). The evolutionary model classifies HCG~31 to be in phase~2 of evolution and the remaining three groups, HCGs~44, 68, and 92, in phase~3.

\subsection{Observations and Data Reduction \label{sec:data_analysis}}

We obtained 21cm \HI maps with the GBT under program GBT07A-093. 
The motivation behind these observations was to probe the \HI distribution in these selected HCGs beyond the region surveyed by B10 and to explore the spatial distribution of the diffuse gas detected by B10. Our objective was to ascertain if the \HI in the diffuse phase has similar spatial distribution as the high-surface brightness (HSB) gas detected in the VLA maps.
We conducted the observations in pointing map mode with 25 individual pointings on a 5x5 grid with a spacing of 4$^{\prime}$ between pointings. The total surveyed area was 25$^\prime$x25$^\prime$. The positions of the centers of the maps are presented in Table~\ref{tbl-sample} and the positions of the GBT pointings are shown in  Figure~\ref{fig-vla_maps} as gray circles.
We employed standard GBT observing routines for the mapping and spent a fifth of the time on the reference position. The reference position was chosen on a blank part of the sky at an offset of 1 degree from the edge of the map in both Right Ascension (RA) and Declination.

To ensure positional accuracy we performed local pointing corrections using the observing procedure AutoPeak. Hence, the error in pointing should be less than GBT's global pointing accuracy of 3$^{\prime\prime}$. This is less that 0.6\% of the full width at half maximum (FWHM) of 9.1$^{\prime}$ for the GBT beam at 1.4~GHz.
The flux was calibrated using L-band calibrators 3C~48 (16.5 Jy), and 3C~286 (15.0 Jy) and the antenna gain was derived for each session individually. We  state a calibration error of $\approx$ 10\% (3$\sigma$) in all the measurements presented in the paper. 

The spectroscopic data were acquired in dual polarization L-band and cover the frequency range between 1.15 and 1.73 Gigahertz (GHz). Using 9-level sampling and two IF settings, each with 12.5 MHz total bandwidth, the spectrometer gives 8192 channels at 1.5~kHz (0.3~km s$^{-1}$) resolution, covering a total velocity range of 2500~km s$^{-1}$. Data were recorded every 10 seconds in order to isolate radio frequency interference (RFI) and to remove the corrupted integrations during data processing. Integration times per pointing, the range of velocities over which \HI\ was detected, and range of errors in \HI\ masses are provided in Table~\ref{tbl-observations}.

Data for two more groups (HCGs~15 and 100) were also obtained under the same program. Unfortunately, they were severely corrupted by extensive RFI. Therefore, these two groups will not be discussed any further.

The data were analyzed using the interactive software package GBTIDL, specifically designed for reduction and analysis of spectral line data obtained with the GBT. Composite spectra were obtained by combining each RFI-free integration using GBTIDL routines and the final spectra were fitted with a low-order  polynomial (second or third order in most cases) for estimating the baseline. 
The good bandwidth stability has allowed us to fit a lower order polynomial and thus made the data sensitive to broad faint spectral features. In addition we chose a large range in velocity to fit for the baseline and hence we expect to detect individual features that are narrower than $\sim$1500~\kms.

The reduced spectra are presented in Figure~\ref{fig-data} in grid format. The position of the grid is marked on the top left corner. The labels A-E mark the decrease in RA and the numbers 1-5 mark the decrease in Declination. Adjacent pointings are not independent due to the large beam size of the GBT (FWHM=9.1$^{\prime}$). Table~\ref{tbl-rho_grid} presents distances between pointings with respect to the central pointing i.e. the center of the map. 
{\em We provide figures showing the GBT spectra overlaid on the VLA maps as additional images in the journal version.}

\section{RESULTS AND DISCUSSION \label{Sec:discussion}}

Atomic gas was detected in all four groups, although their properties vary widely. Here, we interpret out data in conjunction with the GBT spectra for these groups from our earlier study (B10) and VLA imaging of these groups (see Table~\ref{tbl-observations} for references). In our earlier study, we found that the GBT spectra had similar spectral shapes as those of the VLA for most of the groups, although with higher amplitudes. The only additional features detected in the GBT spectra that were not seen in the VLA data were \HI ``wings" that connected distinct velocity components with fainter emission (e.g. HCG~92).

In this study, we focus on the variation of spectral profiles as a function of distance from the center of the group. Our GBT map covers four times the group size (see Table~\ref{tbl-sample}) for HCGs 31 and 92 (where extensive \HI tidal structures were observed in VLA images) and double the group size for HCG~68. HCG~44 is the only group where the mapped area is not significantly larger than the group. For this group, we refer the reader to an extended interferometric map  of the region surrounding the group \citep{serra13}.

\subsection{Spectral Signature of Atomic Hydrogen  \label{Analysis:spec}}

The two ISM-dominated groups, HCGs~44 and 68, show distinct \HI features at the systemic velocities of the member galaxies. For example, HCG~68 shows a double-horned profile at the systemic redshift of HCG~68C (NGC 5350) peaking at the pointing closest to its spatial location. Similarly, the spectra for HCG~44 show distinct features at the optical redshifts of individual members HCG~44A (NGC~ 3190), 44C (NGC~ 3185),  44D (NGC~ 3187) and SDSS~J1017 \citep{serra13}. The spectral regions corresponding to the member galaxies are labeled in Figure~\ref{fig-data}. 

The tidal debris-dominated groups, HCGs~31 and 92, do not exhibit distinct \HI\ features corresponding to the spatial locations and redshifts of the member galaxies. 
HCG~31 exhibits a single component \HI\ feature ($\Delta\rm V_{flux>0} \le 350~km~s^{-1}$ and $\sigma =\rm  57~km~s^{-1}$) encompassing the redshift range of all the group members. \HI\ emission in HCG~31 peaks at the center of the map.
The shape of the profile remains the same with a slight shift of less than 50~\kms in the peak as we move from north to south. There is no change in velocity of the peak from east to west.
Towards the edge of the map in pointings A1 and A2, another strong emission feature was detected between 3170-3600 \kms. There is no corresponding galaxy found in the vicinity of this detection in the Sloan Digital Sky Survey (SDSS) image. This feature is dynamically not connected to the \HI associated with HCG~31 and is separated by more than 300~$\rm km~s^{-1}$. In addition, it was only detected in pointings A1 and A2, and not in B1, B2, and A3. This suggests that the spatial location of this \HI structure is to the Northeast of HCG~31, outside our 25$^\prime$x25$^\prime$ map. It is possible that this feature is not an astronomical source, but was generated by RFI. However, we do not see a detectable variation in its strength between integrations as well as the two polarizations.  Observations from a separate epoch are required to confirm this detection.

The spectra towards HCG~92 show three features. The strongest feature covers a velocity range of 6350-6750~\kms and peaks at position C2. 
This matches the velocity range of some of its member galaxies: HCG~92E and 92D, and extends as far out to the redshift of 92C.  
On the low velocity end, the \HI feature has a large \HI\ ``wing" first detected in our previous study (B10) that can be seen extending out to the second \HI feature in the higher signal to noise spectrum. The second feature lies between  5950-6150 \kms. It peaks at the redshift of galaxy HCG~92F and in positions C2, C3, B2, and B3. 
The third feature is seen between velocities 5600-5800~\kms, thus covering the redshift of HCG~92B and peaks in position B3.

\begin{figure*}
\figurenum{3}
\begin{tabular}{c c} 
\includegraphics[scale=0.285,angle=0]{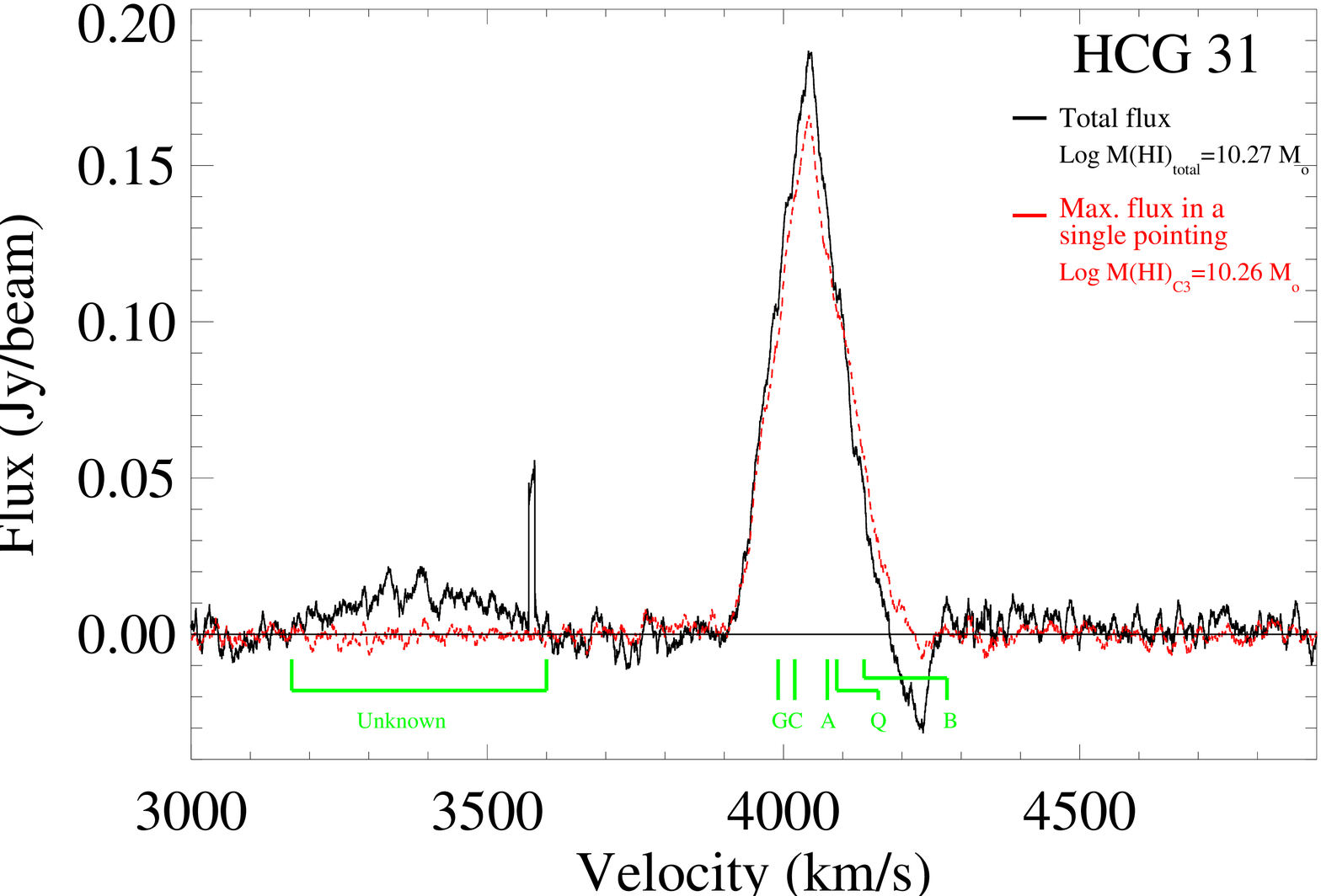} &
\includegraphics[scale=0.285,angle=0]{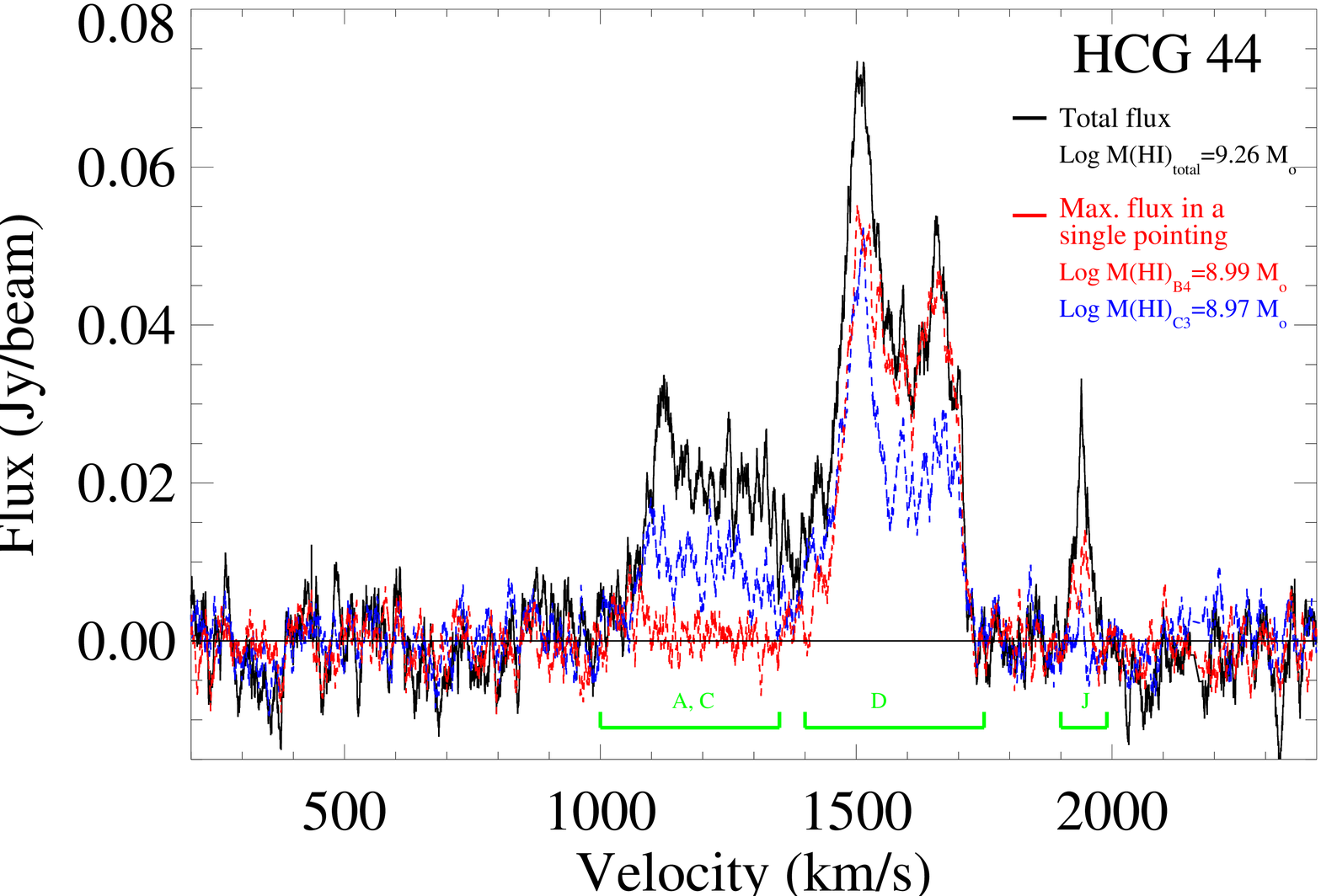} \\
\includegraphics[scale=0.285,angle=0]{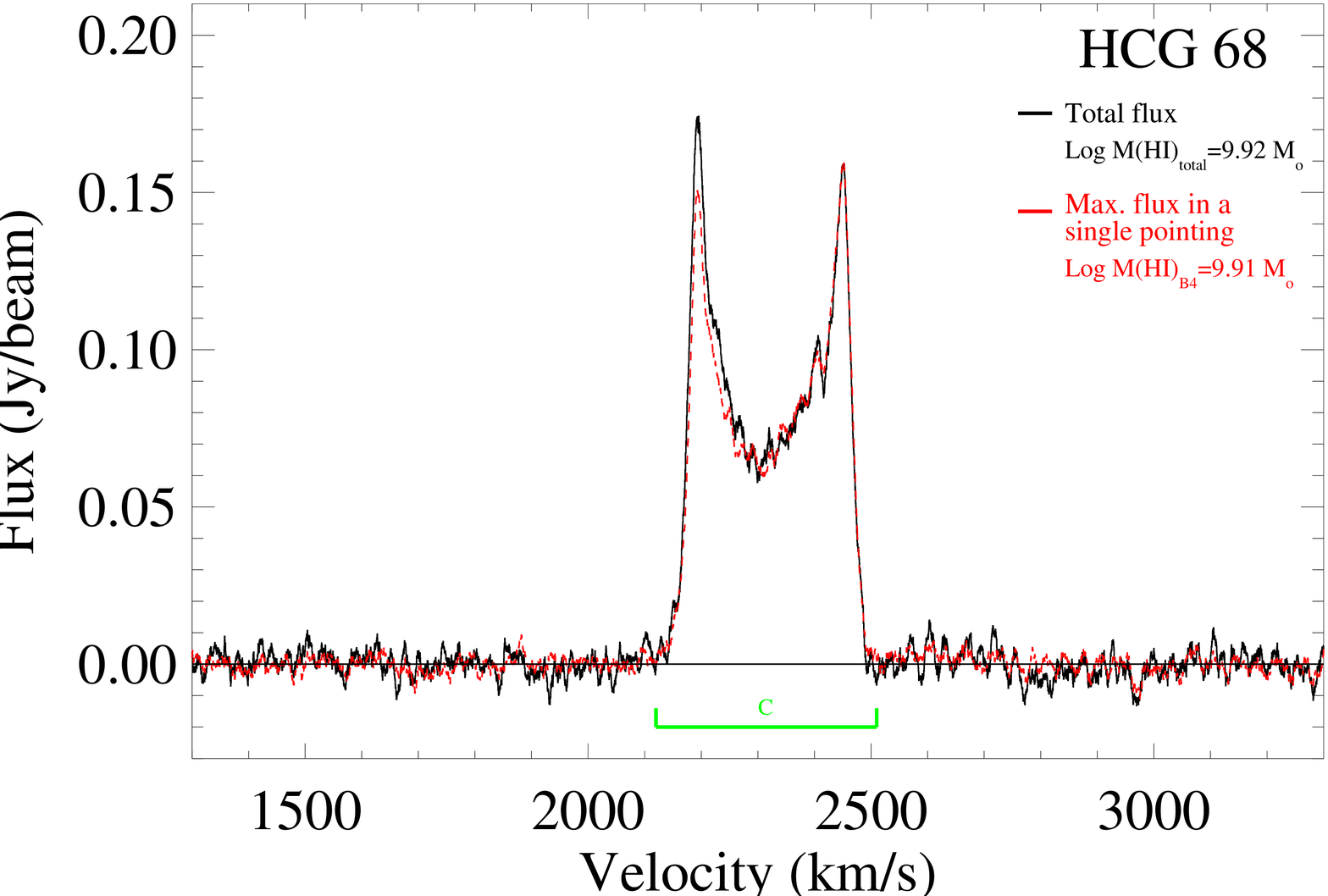} &
\includegraphics[scale=0.285,angle=0]{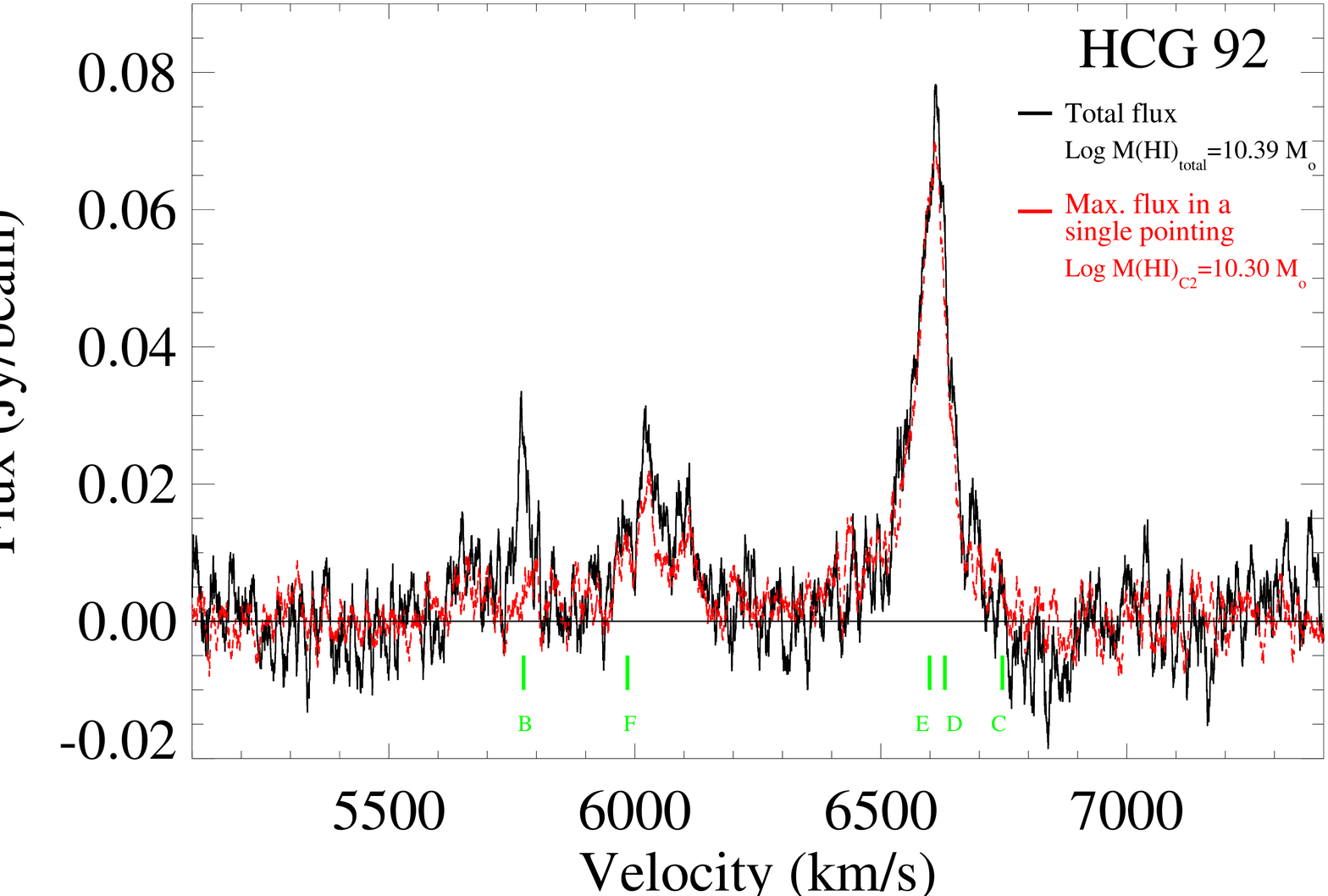} \\
\end{tabular}
\caption{Integrated \textsc{Hi} spectra of all the pointings around the four groups are shown in black (see \S~3.2). For comparison we show the spectrum from the pointing with most HI flux in red. For HCG~44, the spectrum from the central pointing is shown in blue and is of similar strength as the pointing of maximum flux. The labels on the top right hand corner specify the group, its total HI mass, and the HI  mass associated with the pointing containing the most flux.  The peak flux is $98\%$ of the total flux observed in HCGs~31 and 68. This implies that most of the gas is in the central region and gas in the outskirts is just $2\%$ of the total HI in these groups. In HCG~92 the gas is more spread out and the peak is $81\%$ of the total HI whereas in HCG~44 the peak flux is just $54\%$ of the total flux. The broad feature in the spectra of HCG~31 between 3170-3600~\kms is not associated with any optical galaxy in the SDSS images and are certainly not asscociated with the group. Hence that feature was left out of the mass estimated printed on the top right-hand corner. \label{fig-hi_sum}}
\end{figure*}

All the groups with the exception of HCG~68 show blending between various \HI components. 
Most of the \HI in HCG~44 exists in stable disk-like structures (double-horned profiles). The proximity of the systemic velocity between the member galaxies within the groups does result in blending in some of the pointings. However, we found the spatial variation among spectral features to be strongest for the ISM-dominated groups. 
The \HI kinematics in the tidal-debris dominated groups are much more complex. There are no double-horn disk-like signatures in HCGs~31 and 92 (except for 92F). In fact, most of the \HI in both the groups exists as a single peaked profile covering the redshift of all/some the member galaxies. 
The tidally dominated groups shows stronger kinematic blending between various components in all pointings. 
Both these observations are in agreement with the evolutionary model which predicts that in the final stage of evolution (phase3b), \HI , if surviving outside galaxies, would form a single cloud enveloping the entire group with a single velocity profile \citep{VM01}.

 \begin{figure}
 \figurenum{4}
\includegraphics[trim=9mm 0mm 0mm 0mm, clip=true, scale=.53,angle=-0]{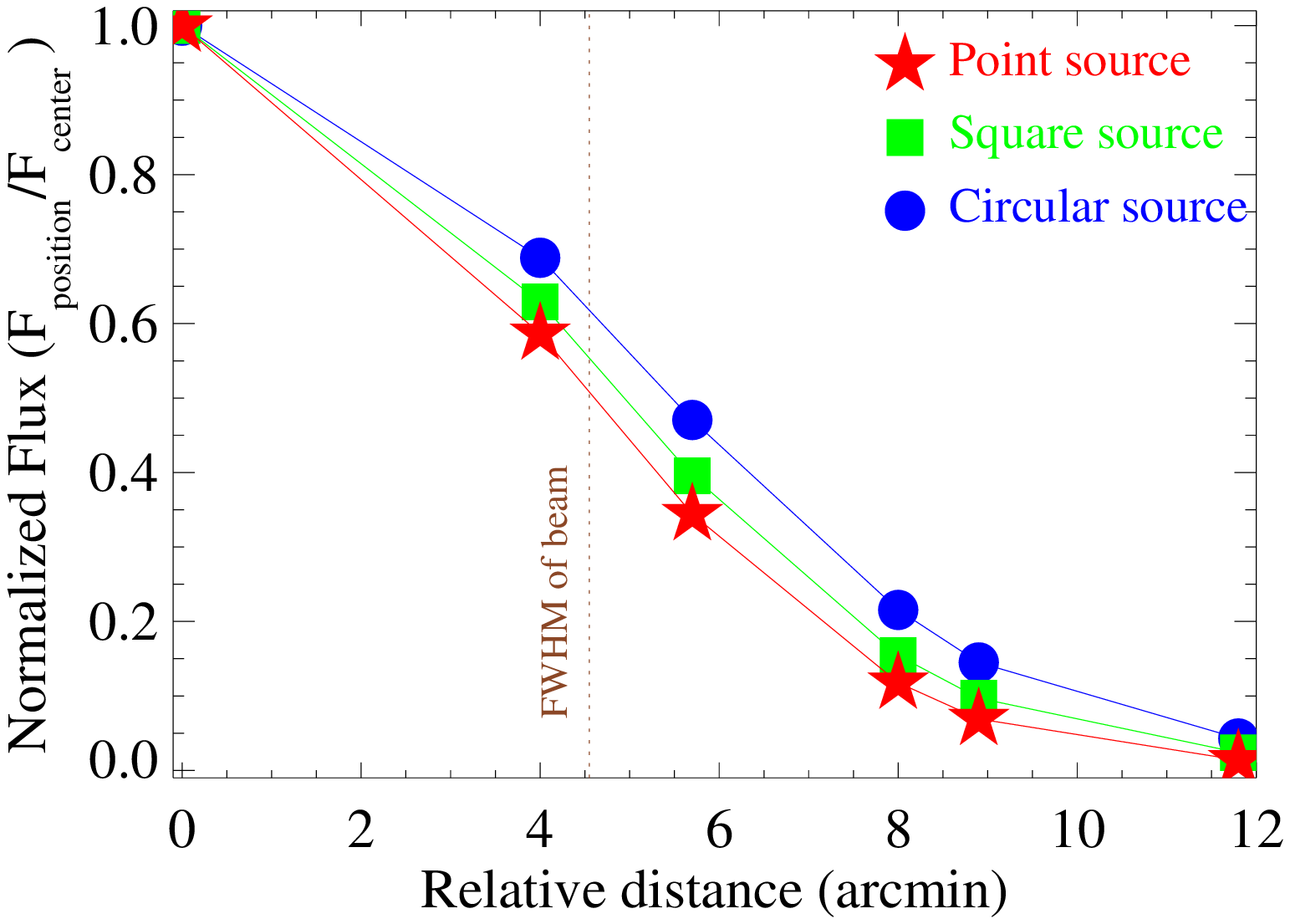} 
\caption{Models illustrating the variation in observed flux (and mass) as a function of distance between the source and the pointing center when observed with the GBT at 1.4~GHz. The GBT beam is a Gaussian of FWHM of 9.1$^\prime$ and hence is sensitive to flux at $>$3\% level for distances upto 12$^\prime$. Three source geometries are considered - point source (red), a square source of side 5$^\prime$ (green), and a circular source of diameter  9$^\prime$ (blue). The sensitivity of the GBT at large offsers from the beam center is higher for more extended targets as indicated by the blue curve representing the circular source with the larger area among the three geometries considered here. We note that these simulations are for symmeteric sources. Large asymmetries in the source geometry can result in significant deviations in the flux sensitivity profile.  \label{fig-models}}
\end{figure}

\subsection{Mass of the Atomic Hydrogen \label{Analysis:mass}}

The atomic gas mass for each of the pointings is presented in Table~\ref{tbl-hi_mass}. The masses are also printed on the top right corner of each of the panels in Figure~\ref{fig-data}. These measurements  were made using the relationship M(HI) = 2.36$\times10^5D^2(S\Delta V)~M_\odot$, where $D$ is the luminosity distance to the galaxy in megaparsecs (Mpc) (column ~4 in Table~\ref{tbl-sample}) and $S\Delta V$ is the velocity integrated \HI\ flux density in Jy~\kms\ .  We also estimated the total \HI\ flux spread over our 5x5 grid by adding spectra from each of the pointings and then normalizing it by a factor of  $\rm (\pi~D^2_{FWHM}/4~ln2)/\it{l}\rm^2=5.86$, where $\rm D_{FWHM}=9.1^{\prime}$ is the FWHM of the GBT beam and $l\rm =4.0^{\prime}$ is the length of each grid on our map.  The resulting spectra are shown in Figure~\ref{fig-hi_sum} and the associated \HI masses are presented in the last row of Table~\ref{tbl-hi_mass}. 

The four groups cover a wide range of \HI masses. Within the surveyed area, the tidal debris-dominated groups HCG~31 and 92 have $\rm 1.86\times 10^{10}$ and $\rm 2.45 \times 10^{10}~M_{\odot}$ of \HI mass respectively  whereas the ISM-dominated groups, HCG~44 and 68,  contain significantly less atomic gas - $\rm 1.82\times 10^{9}$ and $\rm 8.32\times 10^{9}~M_{\odot}$ respectively. These integrated \HI masses are close to the total \HI associated with the groups except in the case of HCG~44 where the \HI extends beyond the GBT mapped area. The total \HI mass of the HCG~44 as reported by \citet[][WSRT imaging]{serra13} was $\rm 2.2\times 10^{9}~M_{\odot}$. 

 On the other hand, the \HI mass detected in the VLA imaging towards HCG92 is merely 1.47 $\rm \times10^{10}~M_\odot$ \citep[][ including Anon2 and corrected for cosmological parameters]{williams02}. 
 This indicates that our GBT measurements found 65\% more \HI in this group than that observed in the VLA image. 
 Similarly, for HCG 68 the GBT detected H I mass is 10\% larger than that detected in the VLA map, 8.32$\rm \times10^{9}~M_\odot$ \citep{tang08}. 
 Both GBT and VLA confirm that the \HI in HCG 68 is contained in the double-horn profile associated with galaxy 68C. 
 Also, in HCG 31 the GBT detected 60\% more H I than that detected in CnB configuration VLA image of 1.126$\rm \times10^{10}~M_\odot$  \citep[][cosmological parameters corrected]{VM05}. 
 The new GBT measurements also match that of single dish measurements by B10 and \citet{williams87} with a difference of less than 5\%. However, the DnC configuration image of HCG 31 was reported to have a larger total H I mass \citep{VM05}. Part of the reason for the difference could be the result of the serious solar interference that effected the DnC configuration data extensively.

For comparison, we plot the pointing with the maximum flux (see Figure~\ref{fig-data}) in red in Figure~\ref{fig-hi_sum}. In all the groups the total flux is larger than the maximum seen in an individual pointing. The difference in flux between the total and the peak measurement ($\rm (M(HI)_{total}-M(HI)_{max})/ M(HI)_{total} $) is about 2\% for HCGs~31, and 68; 19\% for HCG~92; and 46\% for HCG~44. In all the groups the gas on the edges ($\rm M(HI)_{edges}$) is dynamically similar to the central pointing. This is indicative of large spatial coverage of the GBT beam (well beyond its FWHM of 9.1$^\prime$).
The total \HI spectra of HCGs~92 and 44 also show additional \HI features that were not detected in the pointing with maximum flux. These features add substantial amount of \HI to the total atomic gas mass of the groups. Interestingly, the third emission feature in HCG~92 observed between 5700-5800~\kms peaks sharply in the total flux spectrum and not in any of the individual spectra. This is suggestive of a large spatial extent of the gas associated with this feature.

It is noteworthy that while tidal debris-dominated groups contain multiple galaxies, their spectra, when observed with a large beam, is indistinguishable from individual systems. 
This is a valid concern for the blind \HI surveys that do not distinguish between gas associated with individual galaxies and gas in galaxy-groups.
For example, when observed with a beam of FWHM $\rm \approx 15^{\prime}$, the compactness of such groups would make it difficult to identify them as complex \HI tidal structures. However, such \HI-rich groups are very rare both in terms of galaxy distribution as well as in terms of their \HI masses. 
The two tidal debris-dominated groups would exist at the higher-mass end of the \HI\ mass function obtained by \citet{zwaan05} for the nearby universe based on the \HI Parkes All Sky Survey \citep[HIPASS;]{meyer04}\footnote{average gridded HIPASS beam is $\approx$15.5$^\prime$.}. In the case of HCG~92, only 2\% of the extragalactic \HI detections have mass greater than that of this group. On the other hand, the ISM-dominated groups lie close to the peak of the \HI mass function, i.e. $\rm 10^{9.5}~M_\odot$.  The above comparison illustrates the gas richness of these systems.

\subsection{Spatial Extent of the Atomic Hydrogen in HCGs \label{Analysis:spatial_ext}}

\begin{figure*}
\figurenum{5}
\includegraphics[trim=9mm 0mm 0mm 0mm, clip=true, scale=0.52,angle=-0]{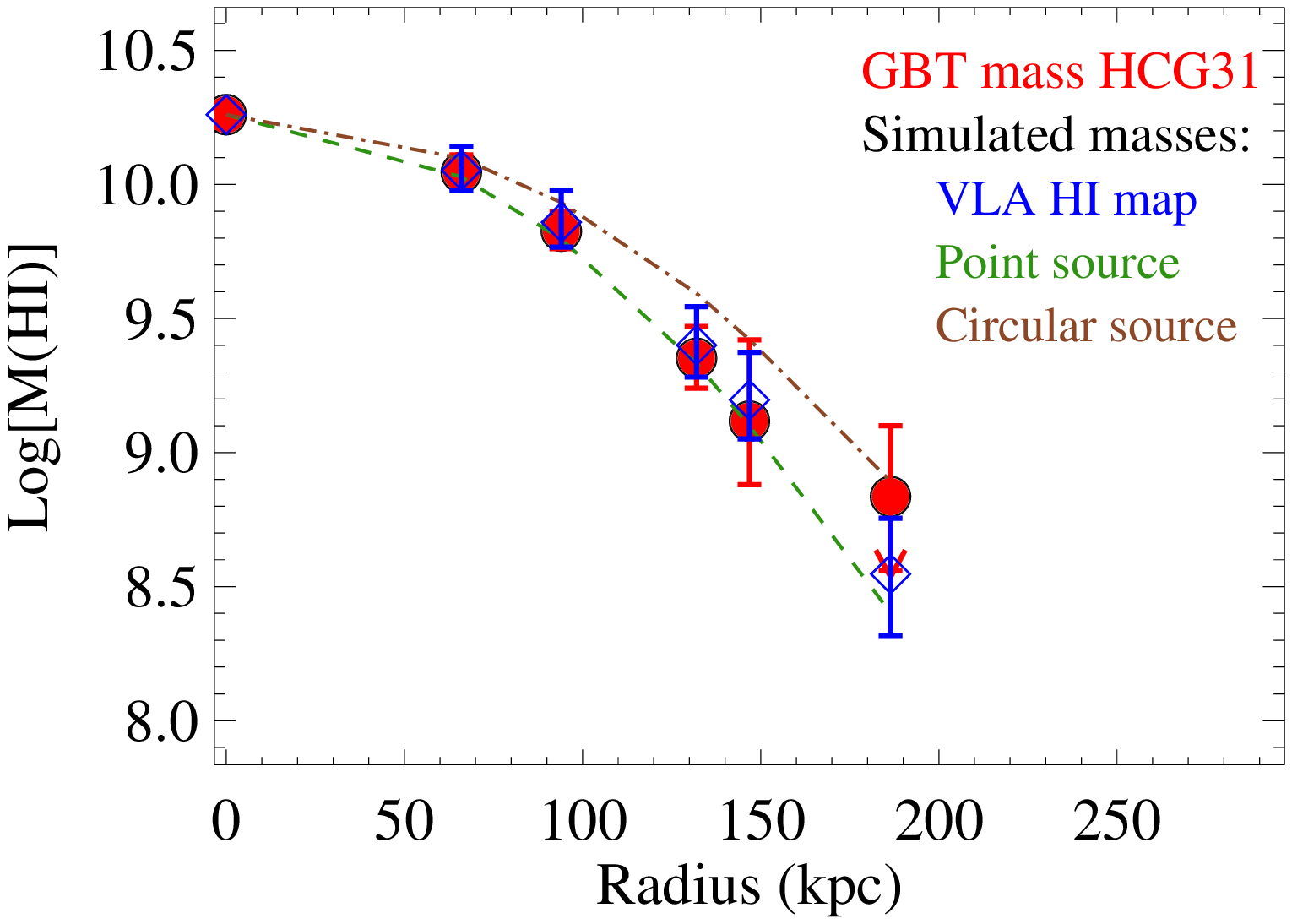} 
\includegraphics[trim=9mm 0mm 0mm 0mm, clip=true, scale=0.52,angle=-0]{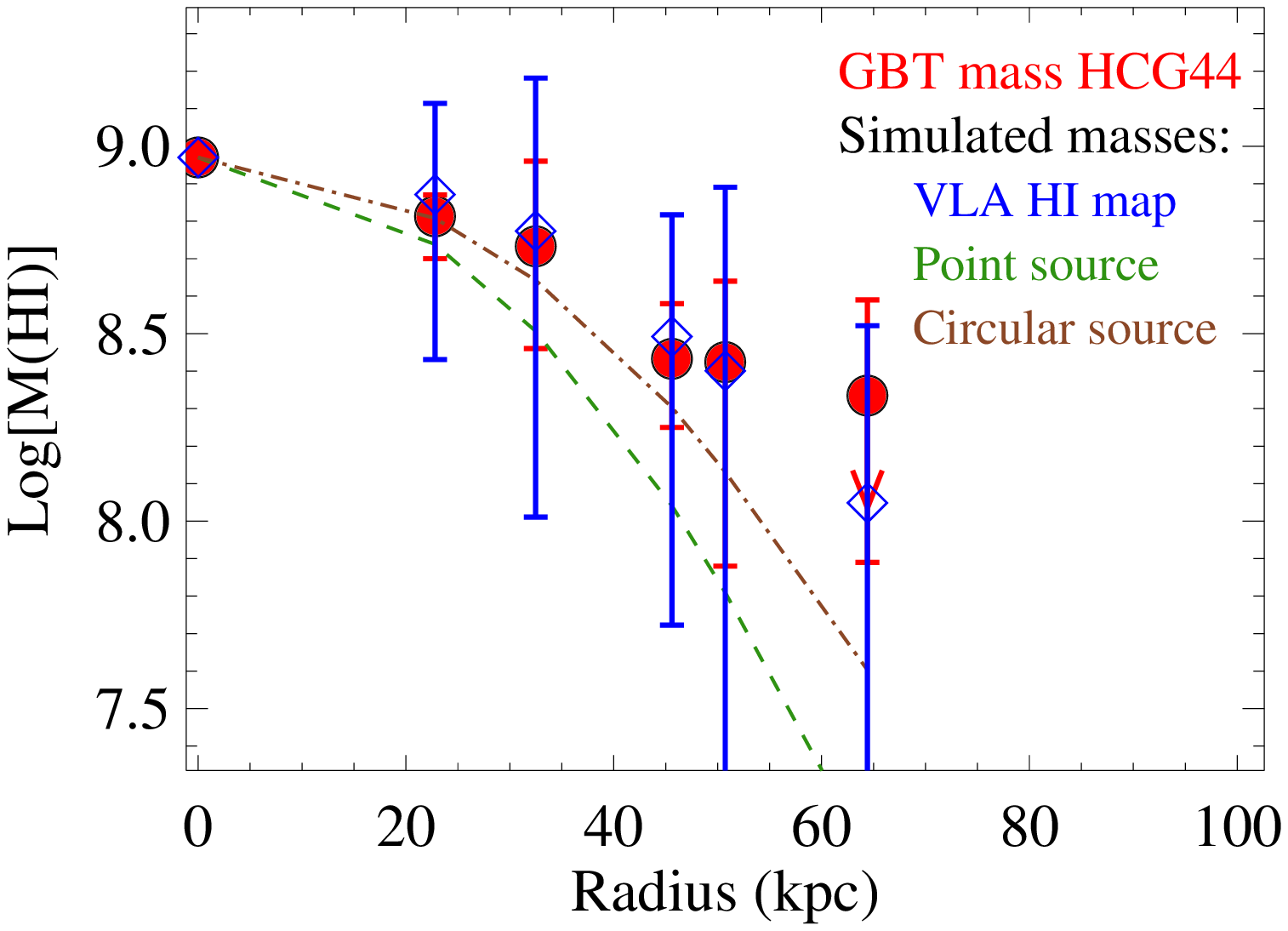} \\
\includegraphics[trim=9mm 0mm 0mm 0mm, clip=true, scale=0.52,angle=-0]{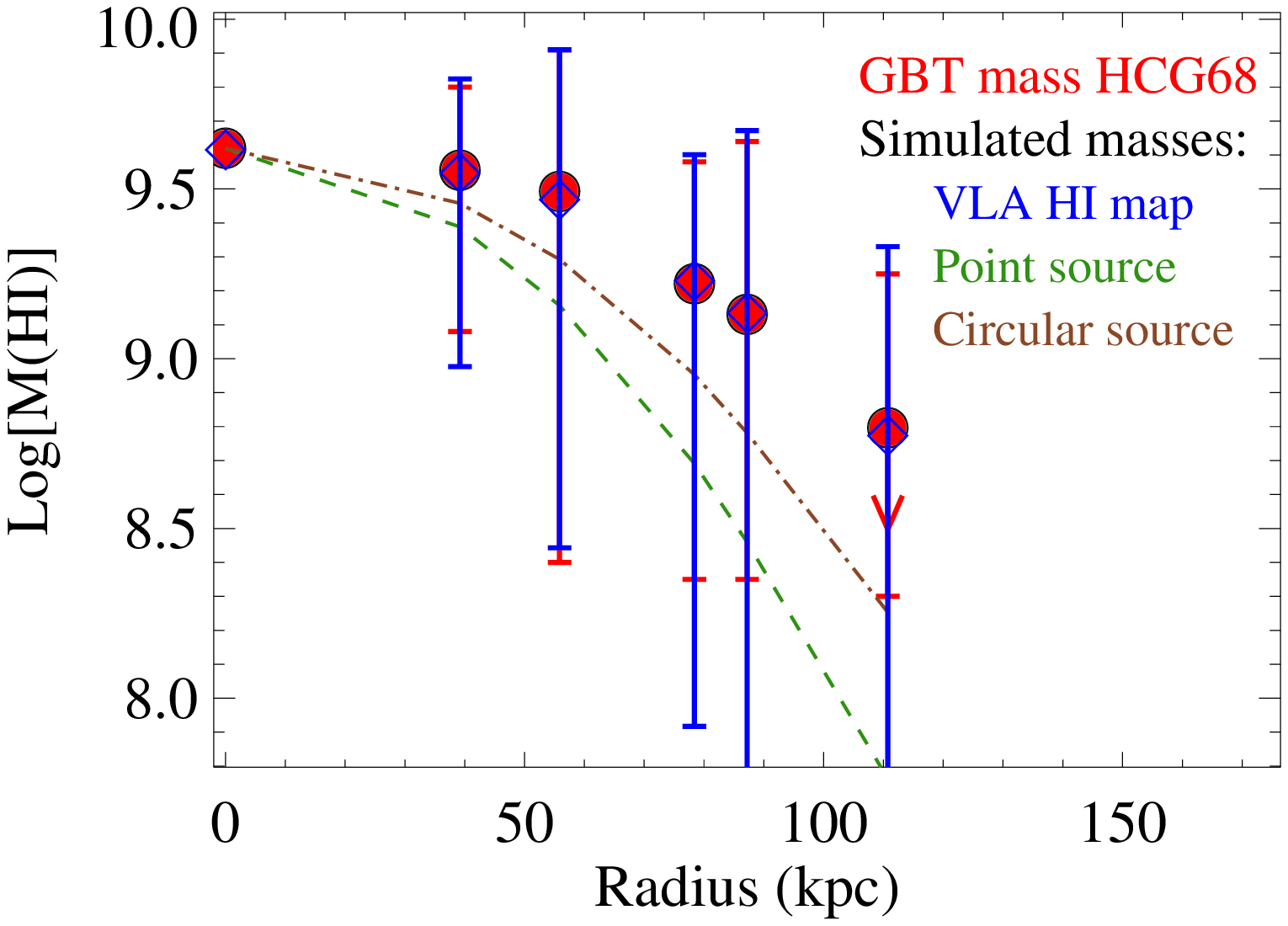} 
\includegraphics[trim=9mm 0mm 0mm 0mm, clip=true, scale=0.52,angle=-0]{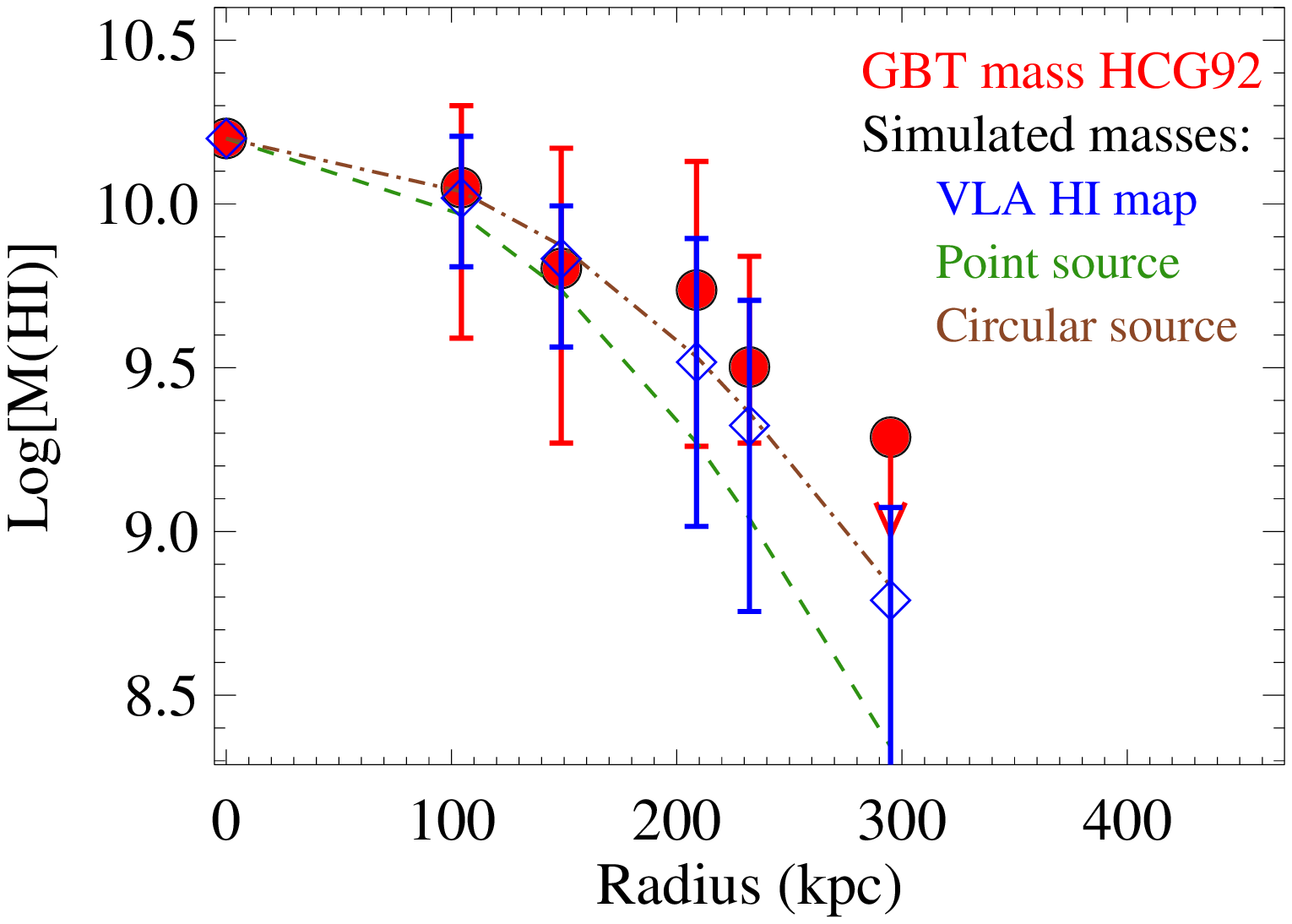} 
\caption{Radial distributions of \textsc{Hi} in the four groups The solid red circles represent the observed \HI mass and the blue open diamonds represent the modeled masses. The ranges in the observed and modeled masses are shown as vertical bars. Each of these values was obtained by averaging 4-8 pointings that are equidistant from the center of the map. The exception to this is the value at zero which represents the central pointing. The models for a point source and a circular source as shown in Figure~ \ref{fig-models} are overplotted in green and brown respectively. It is worth noting that not all the pointings that were averaged are independent, although the modeled masses takes that into account. The model data were normalized at the central pointing (C3) corresponding to zero in the absissa. The data from HCG~31 matches that of a point source whereas the other sources are slightly resolved.  The range of values in the observed and the modeled data match well for HCG~31 and 68 whereas HCG~44 and 92 show large differences in the spread.  \label{fig-h31_radial}}  
\end{figure*}

Our data consist of 25 spatially distinct pointings. However, they are not independent. 
This is because the pointings are separated by 4$^\prime$ whereas the GBT beam is a two-dimensional Gaussian of FWHM $\rm =9.1^{\prime}$ \citep{boothroyd11}.  As a result, the GBT beam is sensitive to sources at  $\sim \rm 5\%$ levels at even 11.5$^\prime$ away from the pointing center. Figure~\ref{fig-models} illustrates the effect of the GBT beam shape on its sensitivity as a function of distance between the source and the pointing center. 
To further illustrate the effects of source geometry on the sensitivity of the GBT, we also considered three source  geometries - point source, source in the shape of a square of side 5$^\prime$, and a circular source of diameter 9.1$^\prime$.
 The sampling in these simulations matches that from our observations. These simulations imply that none of our pointings are completely independent of the central pointing. This also suggests that the strength of the flux observed at any position is dependent on the shape/morphology of the source.

A comparison of our data with the models shown in Figure~\ref{fig-models} are presented in Figure~\ref{fig-h31_radial}. The plots show the radial distribution of \HI as a function of distance from the central pointing in the units of physical distance. The observed masses are shown in red circles. The modeled \HI masses are for two test cases: a point source and a circular source of diameter 9.1$^\prime$are shown as a green dashed line and a brown dotted dashed line, respectively. Unresolved systems are close to the green dashed line and barely resolved systems are sightly above the brown dotted-dashed line.  The \HI structures in HCG44 and HCG~68 are resolved whereas in HCG~31 they are  unresolved. Also, \HI structures in HCG~92 are partly resolved.  We also show the models based on the \HI distribution seen in the VLA maps (discussed later) as blue diamonds.

To disentangle the effects of the beam, we imaged the data using the GBT Mapping Pipeline. However, the sampling of data, i.e. 25 pointings, was found to be significantly lower  for studying distribution and morphology of the gas. We also tried to extracted spatial information from the data  using the Richardson-Lucy algorithm to deconvolve the GBT beam pattern. Unfortunately, the results were found to be of low spatial resolution and consistent with those found with the GBT Mapping Pipeline.

In the absence of high resolution maps of the groups with the GBT, what can we learn about the \HI distribution in these groups? It turns out that although we cannot image the \HI, we can still ask whether the GBT data are consistent with the spatial distribution of HSB \HI seen by the VLA. For instance, if the excess gas that was missed by the VLA but detected by the GBT (B10), has a similar spatial distribution then we expect to see the GBT data agree with the VLA data. Alternatively, if the excess gas has a very different spatial distribution, then the GBT and the VLA maps should be inconsistent. 

We simulated the expected masses for each of our GBT pointings using the VLA \HI distribution as a template for the source structure. The hypothesis that we  are testing here is that the total \HI (faint $+$HSB) distribution traced by the GBT has a similar spatial distribution to the HSB features detected by the VLA. 
Since the GBT is sensitive to more diffuse gas in addition to the HSB features detected by the VLA, we normalized the modeled masses to one of the pointings. In this case, we used the pointing with the maximum observed mass. This is an arbitrary choice and normalization can be done at any pointing. The normalization process assumes  a ``constant" value of  the ratio total-to-HSB \HI for the entire group. So in this case the ``constant" for normalization was chosen to be the ratio at the pointing of peak \HI flux/mass. 

 A good match between the observed and modeled masses would mean that the ratio is constant across the mapped region. In other words, this would imply that the total (faint+HSB) \HI distribution traced by the GBT has a similar spatial distribution as the HSB features detected by the VLA. Conversely, any deviation would indicate that the assumed ratio is not constant spatially, thus implying a different spatial distribution of the faint gas.
It is worth noting that since the modeled masses were normalized, the deviation can be both positive and negative.
A deficit in observed mass does not correspond to the missing \HI in the GBT observations. Instead it simply corresponds to a smaller total-to-HSB \HI ratio (also equivalent to smaller faint-to-HSB \HI ratio) than that of the pointing with maximum flux. 
 

 Figure~\ref{fig-data} shows the observed masses in black at the top right-hand corner and the modeled masses in color. Deviations between the observed and the modeled masses that exceed more than 0.1~dex ($\equiv \rm \pm 20\%$, which is twice the uncertainty in the measurements) were considered significant and to be beyond the noise in our measurements. The modeled masses for the pointings, where they are in agreement with the observed masses, are printed in green (i.e. $\rm |M_{mod}-M_{obs}| \le 0.1~dex$). Pointings where the modeled masses differ from the observed masses by more than 0.1~dex  are indicated in red (i.e.  $\rm |M_{mod}-M_{obs}| > 0.1~dex$). Pointings where the modeled masses are smaller than the limiting \HI masses (i.e. $\rm M_{mod}\le M_{lim}$) are printed in blue.

\subsubsection{HCG~31}  The observed and the modeled \HI distributions are consistent for most of the pointings for HCG~31. This indicates that most of the excess gas seen in the GBT spectra is associated with HSB features and has a similar spatial distribution. Hence, we conclude that most of the \HI is centrally located and is consistent with being associated with the ISM/tidal structures of member galaxies. The dispersion in the observed masses as a function of radial distance and that of the modeled masses suggest that the VLA and the GBT spatial distribution are similar. In fact, the observed distribution is consistent with a point source (relative to the GBT beam) as seen in Figure~\ref{fig-h31_radial}. 
We have also analyzed the observed \HI distribution by deconvolving the GBT beam from the data using the Richardson-Lucy algorithm and found consistent results.

Pointing E1 shows the maximum difference between the observed and modeled masses corresponding to a mass excess of $\rm M(HI)_{obs}- M(HI)_{sim} ~=~7 \times 10^8~M_{\odot}$. This points to the presence of an \HI structure at the edge of the surveyed region in the Southwest direction such that none of the other pointings have detected it at $>$20\% levels. We also observed small discrepancies between the modeled and observed masses in pointings B1, D5,  and E4. Unlike E1, the observed masses in these pointings were smaller than the modeled masses. 
This indicates that the faint \HI distribution is more skewed in the Northeast to Southwest direction than the HSB features.

\subsubsection{HCG~44} The \HI in this group shows distinct spatial variations corresponding to the positions of the member galaxies. All the \HI\ features peak at the position of their host galaxies and gradually fall off with distance. 
The strongest \HI feature is associated with HCG~44D (NGC~3187) and peaks at position B4. This pointing was used to normalize our modeled masses. Since the VLA map did not cover channels beyond 1800~\kms, the spectral feature between 1900-2000~\kms associated with galaxy SDSS J1012 (marked as `J' on the spectra) was not covered in the VLA maps. The total \HI mass, including the feature corresponding to SDSS J1012, is printed in black at the top right corner in Figure~3. For pointings B5, C4, C5, D4, and D5, we also present the \HI masses for gas at $v<1800$~\kms in grey in parenthesis. This is the most appropriate quantity to compare with the modeled fluxes estimated from the VLA map.

Most of the pointings in the Northwest to Southeast direction show significant excesses from the modeled masses. 
This ratio of total to HSB \HI increases rapidly with distance from B4. The ratio,  $\rm  (M(HI)_{GBT} - M(HI)_{sim})/M(HI)_{sim}$, peaks at pointings E5 ($\sim 700\%$) and C1 ($\sim 600\% $). All the pointings towards the Northeast are in agreement. This suggests that most of the faint gas lies towards the South of B4. This is not surprising as two of the HI-rich neighbors of 44D are South and Southwest of its position.

Figure~\ref{fig-h31_radial} shows the radial distribution of \HI in HCG~44. The normalization was done at the central pointing for ease in comparison.  The profile clearly deviates  from a point source or even circular source filling the  central 9.1$^{\prime}$. \HI associated with this group extended. Interestingly, the \HI radial profile of this group shows significant differences in the dispersion between the observed and modeled masses.

\subsubsection{HCG~68} The \HI\ detected in this group comes from the double-horn profiled centered at the redshift of galaxy HCG~68C (NGC~5350). The spatial position of \HI is consistent with originating in the \HI disk of this blue spiral galaxy HCG~68C. The \HI peaks at position B4 that is closest to the position of galaxy HCG~68C.  The good match between observed and modeled masses thus suggests that the total \HI distribution is almost identical to the HSB features observed by the VLA. 

The \HI profile and distribution are consistent with a large unperturbed \HI disk as imaged by the VLA and shown in Figure~\ref{fig-vla_maps} \citep[also see Figure~25 of][]{tang08}. The profile is highly symmetric and shows no sign of dynamical tidal deformation.  
The \HI diameter of the disk is $\sim$ 9$^\prime$ from the VLA \HI\ maps.  At the redshift of the galaxy, this implies a disk size of $\sim$86~kpc in diameter. It is rare for such an extended disk to show no sign of tidal interaction in a group environment. Therefore, galaxy HCG~68C is most likely a recent addition to the group and could still be infalling.

\subsubsection{HCG~92} The relative strengths of the three features associated with the \HI in HCG~92 show variations with positions suggesting the presence of distinct features at different positions and velocities.
The strongest feature peaks at position C2 suggesting it is offset from the center of the group by more than 2$^\prime$. The velocity range of this features matches that with Arc-N, Arc-S, and NW-HV that were identified in the VLA imaging \citep[see Figure~8 from][]{williams02}. The positions of these structures are consistent with the offsets seen in our GBT data.

Figure~\ref{fig-data} shows the distribution of modeled masses and its comparison to observed masses. The modeled fluxes are normalized at pointing C2. We found that the distribution is consistent in the center of the group, however it varies significantly at the outskirts namely at pointings C1, B1, A2, A3, B4, C4, D4, D3, D2, D1, and E2.  While the first four pointings show an excess in the observed \HI masses, the remaining seven pointings show a deficit. This represents a displacement of the total \HI distribution from the HSB features towards the North of the group.
This is also consistent with the fact that the high-velocity \HI wing between 6300-6530~\kms, first observed by B10, is primarily localized between pointings B2, B3, C2, and C3. This suggests that the gas associated with the \HI ``wing"  is located close to the intersection between these pointings.

\subsection{Nature and Origin of the Diffuse Extended Gas in HCGs\label{Analysis:nature}}

The spatial distribution of the diffuse gas is similar to HSB features in HCGs~68 and 31.  This implies that the diffuse gas is in faint extensions of the HSB gas.
On the other hand, HCGs~92 and 44 shows that most of the diffuse gas is concentrated in a localized region and has an overall distribution different from that of the HSB features.
The discrepancy (excess \& deficits) between observed and the modeled masses suggests that the diffuse gas has a different spatial distribution than the HSB structures.
 The spectral profile of the total \HI being similar to HSB features suggests that the diffuse gas may be associated with older tidal structures. 
For example, HCG~44 shows dynamically disturbed HSB \HI structures near the pointings with the maximum discrepancy at C1 and E5. Similarly, the location of the gas associated with the high-velocity``wings" (flux between 6300-6530 \kms seen only in the GBT data) is consistent with the Arc-N, which exhibits not only complex morphological and dynamical structure, but is also spatially displaced from any of the member galaxies, thus indicating a tidally disrupted structure.

The existence of faint diffuse structures associated with tidal interactions in multi-galaxy interacting systems is well known. One such example is the extended tidal structures seen in the nearby galaxy-group, M81 \citep{yun94,chynoweth08}. Strong dynamical disruption in the spectral profiles of the cold gas in HCGs~44 and 92 indicate their origin to be related to the transformation of the ISM of individual galaxies into the intragroup medium.

\subsection{Fate of the Gas and its Role in the Evolution of the Group }

\begin{figure}
\figurenum{6}
\includegraphics[trim=20mm 95mm 150mm 17mm, clip=true, scale=0.75,angle=-0]{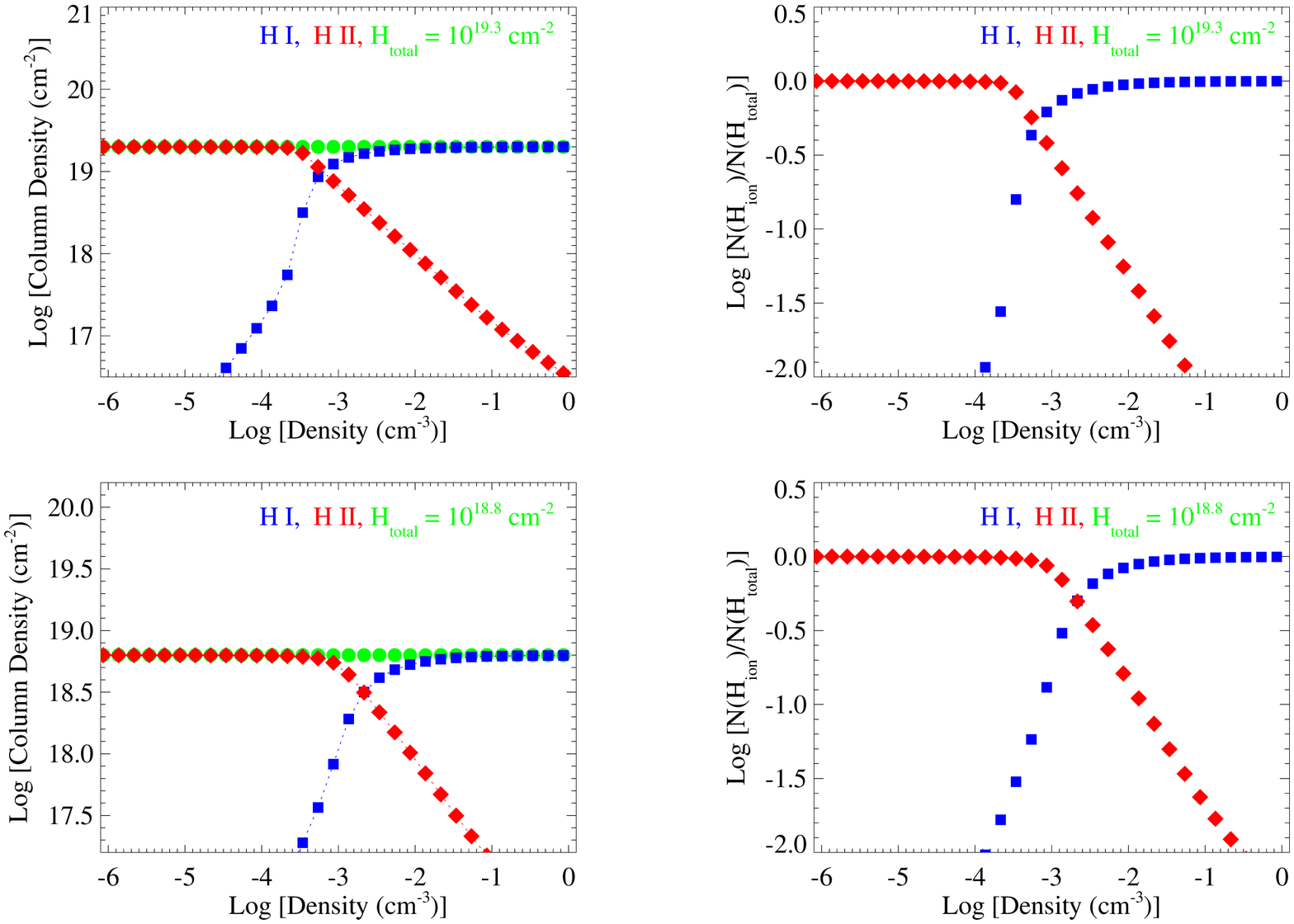} 
\caption{Variation of column density for an \textsc{Hi} cloud as a function of density that is being irradiated the cosmic UV background. The model was generated using photoionization code CLOUDY for a cloud with a hydrogen column density of 2$\times\rm 10^{19}~cm^{-2}$, shown as the green dots. The blue and the red dots refer to neutral and ionized hydrogen as a function of the density of the cloud. As the density drops, the cloud becomes increasingly susceptible to ionization by the cosmic UV background. The ionization state of the cloud changes from primarily neutral to primarily ionized between a density of $\rm 10^{-3}~cm^{-3}$ and $\rm 3\times 10^{-4}~cm^{-3}$. \label{cloudy_hm}}
\end{figure}

In this sub-section, we discuss the survival timescales of diffuse structures against photoionization from the ultraviolet (UV) background and ionizing radiation produced by young star forming regions in the groups.  
The \HI in the ISM remains immune to ionization by the background UV radiation owing to the high densities inside their host galaxies. However, after being stripped, the ISM clouds are unable to maintain their internal pressure and hence expand due to their internal velocity dispersion. As a result the density and the column density of the clouds drop. This can be expressed as a function of time, $t$, and the internal velocity dispersion, $v$, as 
\begin{equation}\label{density_col_density}
 n = n_i~L^3_i/(L_i+ {\it vt})^3 ~{\rm and} ~  N= N_i~L^2_i/(L_i+{\it vt})^2 
\end{equation}
where, $ L_i, ~n_i, ~and ~N_i$ are the initial sizes, densities, and column densities. 
Consequently, the tidally stripped \HI structures expand with time and become more and more susceptible to ionization from the cosmic UV background.

They can also be ionized by extreme ultraviolet  radiation that may be escaping from the star forming regions (if any) in their vicinity. Tidal interactions are known to trigger the formation of young stars. At times some of the starforming regions may be hosted at the tips of extended tidal tails and resemble dwarf galaxies. Such structures have been appropriately termed as tidal-dwarf galaxies \citep[TDG;][]{zwicky56, schweizer78, mirabel92,duc99}. Like other interacting galaxies, HCGs have been found to host multiple star-forming TDGs and individual star clusters outside the parent galaxies \citep[][and others]{hunsberger96, Iglesias01, gallagher01, xu05, torres-flores09, gallagher10, de12, konstantopoulos13, eigenthaler15,fedotov15}. The young stars produce large amounts of ionizing photons and given the right conditions \citep[such as strong stellar winds][]{vogt13, borthakur14} a fraction of the ionizing flux may escape the molecular cloud cocoon where the young stars are embedded. 

\begin{figure*}
\figurenum{7}
\includegraphics[trim=0mm 40mm 0mm 50mm, clip=true, scale=0.420,angle=-0]{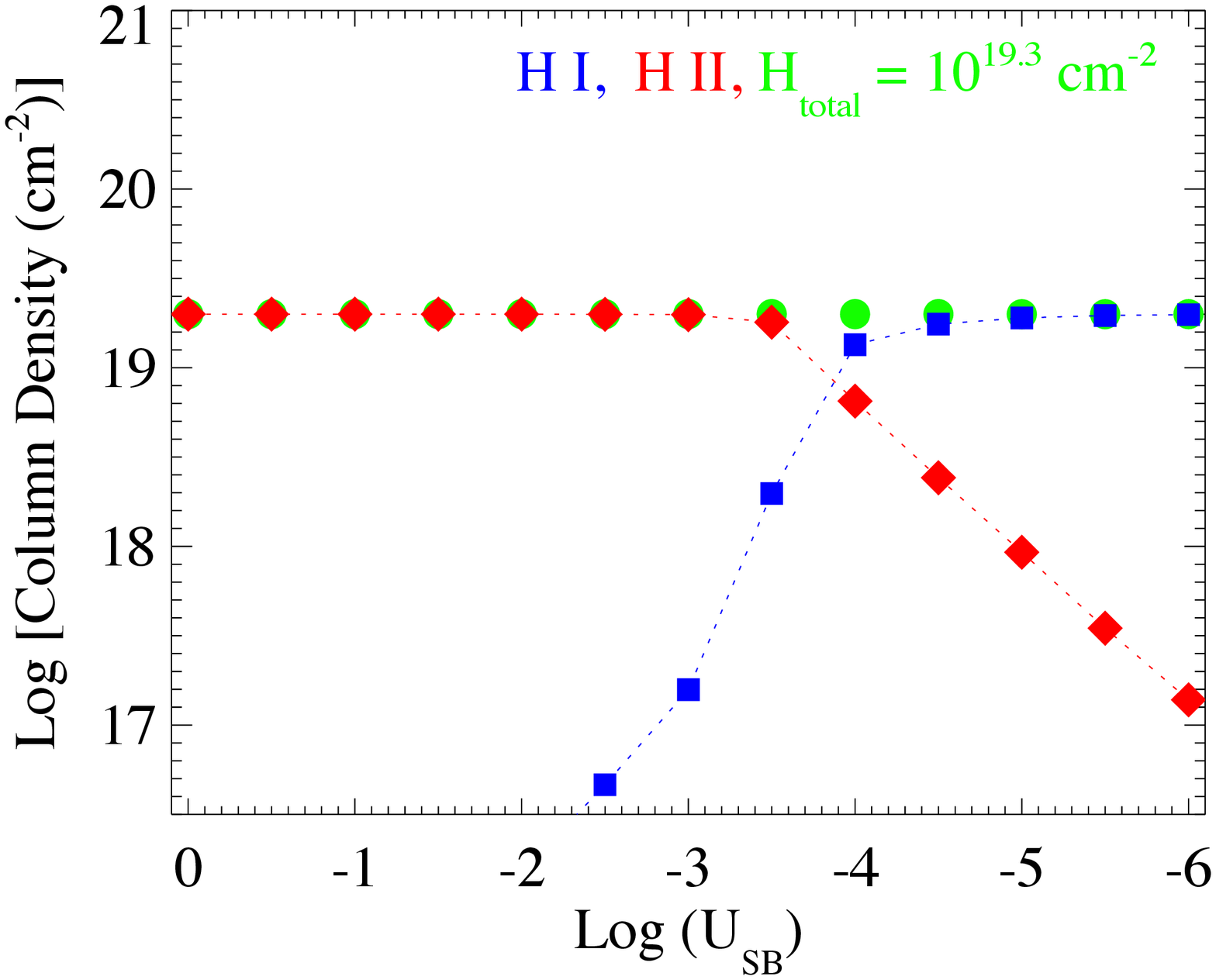} 
\includegraphics[trim=0mm 48mm 0mm 50mm, clip=true, scale=0.420,angle=-0]{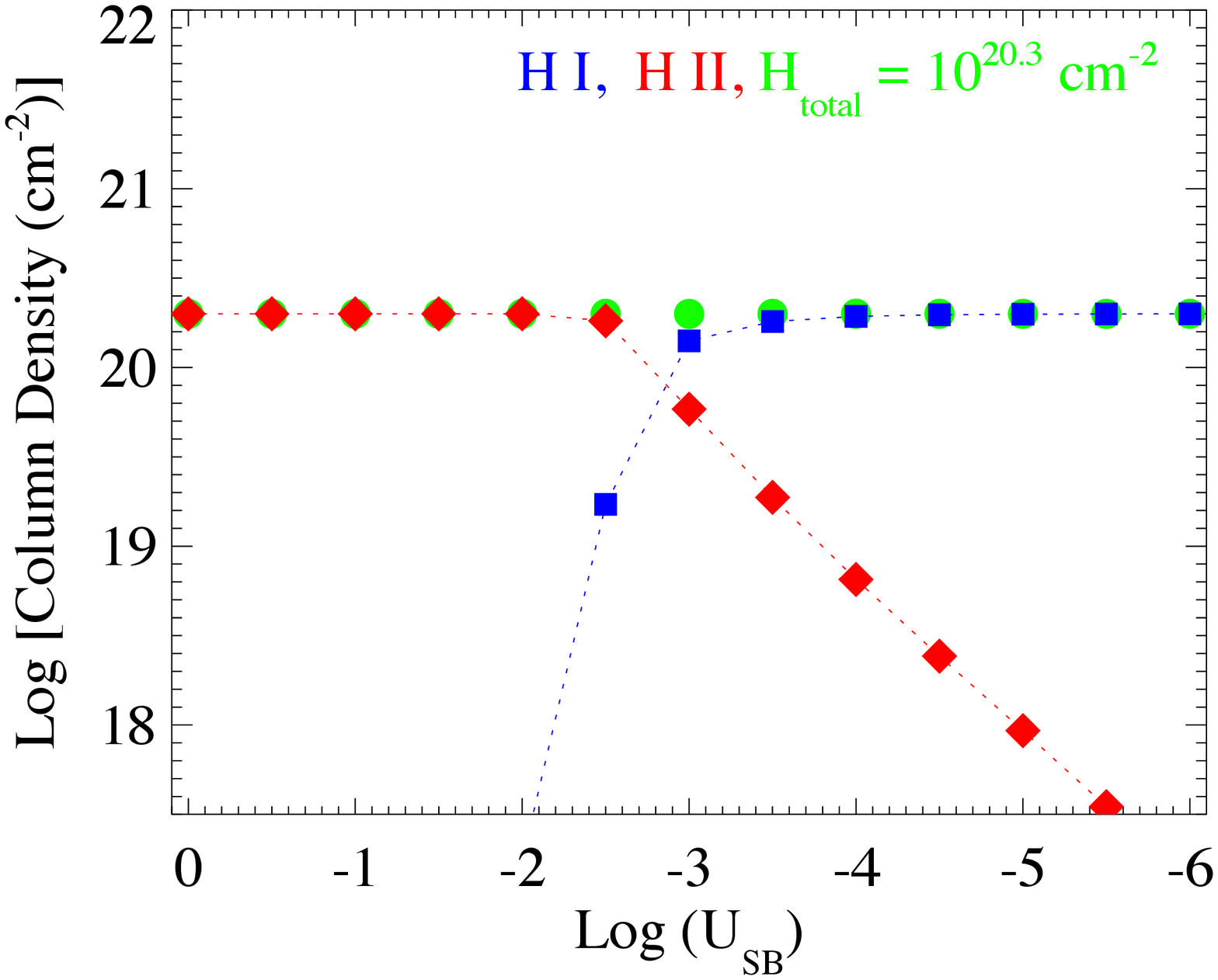} 
\caption{Variation of column density of neutral and ionized hydrogen as a function of ionization parameter, $\rm U_{SB}$ (as defined in Eq. 2), for a cloud. The $\rm U_{SB}$ encompasses cloud properties such as its density and distance from the radiation source. We present two photionzation models that were generated using code CLOUDY. The left panel shows the ionization for a cloud of total hydrogen column density of  2$\times\rm 10^{19}~cm^{-2}$ and the right panel for a column density of 2$\times\rm 10^{20}~cm^{-2}$ respectively.
The total column density is kept constant (shown in green) while the variation of the neutral and ionized components are shown in blue and red respectively.  As the ionization parameter increases so is the ability of the radiation field to ionize the cloud.  As expected the partial ionization is achieve at lower ionization parameter for low column densities as can be seen from left panel as opposed to the right panel. The ionizing source for both the runs is a young stellar population (continuous starburst) of 4~Myrs of age. \label{cloudy_sb}}
\end{figure*}

\subsubsection{Ionization by the Cosmic UV Background}

In order to investigate the lifetime of such tidally stripped neutral gas structures, let us consider photoionization of the neutral gas by the UV background. 
Observations suggest that atomic gas at column densities greater than $\rm 2\times 10^{19}~cm^{-2}$ is immune to ionization from the UV background \citep{corbelli89,vangorkom91}. This result is based on deep 21~cm \HI mapping of the disks of normal galaxies that reveal a sharp cutoff in the disks at this \HI column density.
Motivated by this argument, we can consider the time it would take a stripped gas cloud to reach this column density. This will be its survival time. 
For example, a stripped ISM structure of initial size 1~kpc with density of 1~$\rm cm^{-3}$ expanding at 20\kms would take $\approx$500~Myrs to reach this column density.  At that time, the density will drop by more than three orders of magnitude to $\rm 7.5 \times 10^{-4}~ cm^{-3}$. If it continues to expand beyond this time then the structure will be susceptible to ionization. Hence we can conclude that such a \HI structure will survive for at least 500~Myrs. 
On the other hand, the density of the hot IGrM is expected to be $\rm 4 \times 10^{-3}~-~9 \times 10^{-4}~cm^{-3}$ \citep{freeland08,rasmussen08}. Therefore, pressure equilibrium requires that tidally stripped cooler ISM clouds will cease to expand much before reaching densities comparable to that of the IGrM.  Hence, the clouds would continue to survive for a much longer period of time.

To explore the fate of the diffuse gas further, we performed detailed photoionization modeling using the code CLOUDY \citep{cloudy}, where we confirmed that clouds with neutral column density greater than 2$\times\rm 10^{19}~cm^{-2}$ survive photoionization from the UV background. These simulations used the latest corrections to the UV background i.e., ionization parameter,  U=$\rm \Phi_{HM}/(c~n_H)$, where $\rm \Phi_{HM}= 2750~photons~cm~^{-2}~s^{-1}$ \citep[similar to that used by ][]{borthakur13}, based on the published values by \citet{haardt_madau12} and \citet{giroux_shull97}.   
Even for clouds at $\rm N(H)=2\times 10^{19}~cm^{-2}$, the ionization fraction due to photoionization would depend on the density of the cloud (see Figure~\ref{cloudy_hm}).
The ionization fraction changes drastically for clouds with densities between $\rm 10^{-3.5}~and~10^{-3.0}~cm^{-3}$. 
Clouds with densities greater than $\rm 7.5 \times 10^{-4}~cm^{-2}$ have \ion{H}{1}/$\rm H_{total}$ fraction of more than 50\%. 
In conclusion, as long as the column densities do not drop below this value, the  diffuse \HI structures will be able to survive ionization by the UV background.

It is worth noting that the timescale calculations are dependent on the assumption for initial column density of the gas. Here we have chosen the initial values as the average density and column density expected in the spiral arm of a normal galaxy. However, the higher column density knots will be much more immune to ionization and may produce TDG or star-forming clumps.  Such clumps are also likely to be detected in the VLA imaging due to their higher column densities. These clouds may survive indefinitely if they are not disrupted by other forces such as winds from young star-forming regions or further tidal disruption. Nevertheless, the large extent of the diffuse gas and their fairly long survival times suggest that the TDGs and the diffuse gas may have the same origin. For example, a back of the envelope calculation shows that if a tidal arm is pulled out of the galaxy at the rotational speed of $\sim \rm 200$~\kms for 500~Myrs then it will be displaced by 100~kpc from its host galaxy. Incidentally \citet{demello08} found TDGs in the \HI envelope of HCG~100 at about 100~kpc away from the member galaxies. This indicate that the origin of TDGs and diffuse gas in some of the groups may be via the same process(es).
Furthermore, the diffuse faint gas may eventually get accreted by the any of the galaxies, TDG, or larger star-forming clumps and help in sustaining star-formation in the group. 

\subsubsection{Ionization by Young Stellar Populations}

\begin{figure}
\figurenum{8}
\includegraphics[trim=0mm 05mm 0mm 05mm, clip=true, scale=0.300,angle=-0]{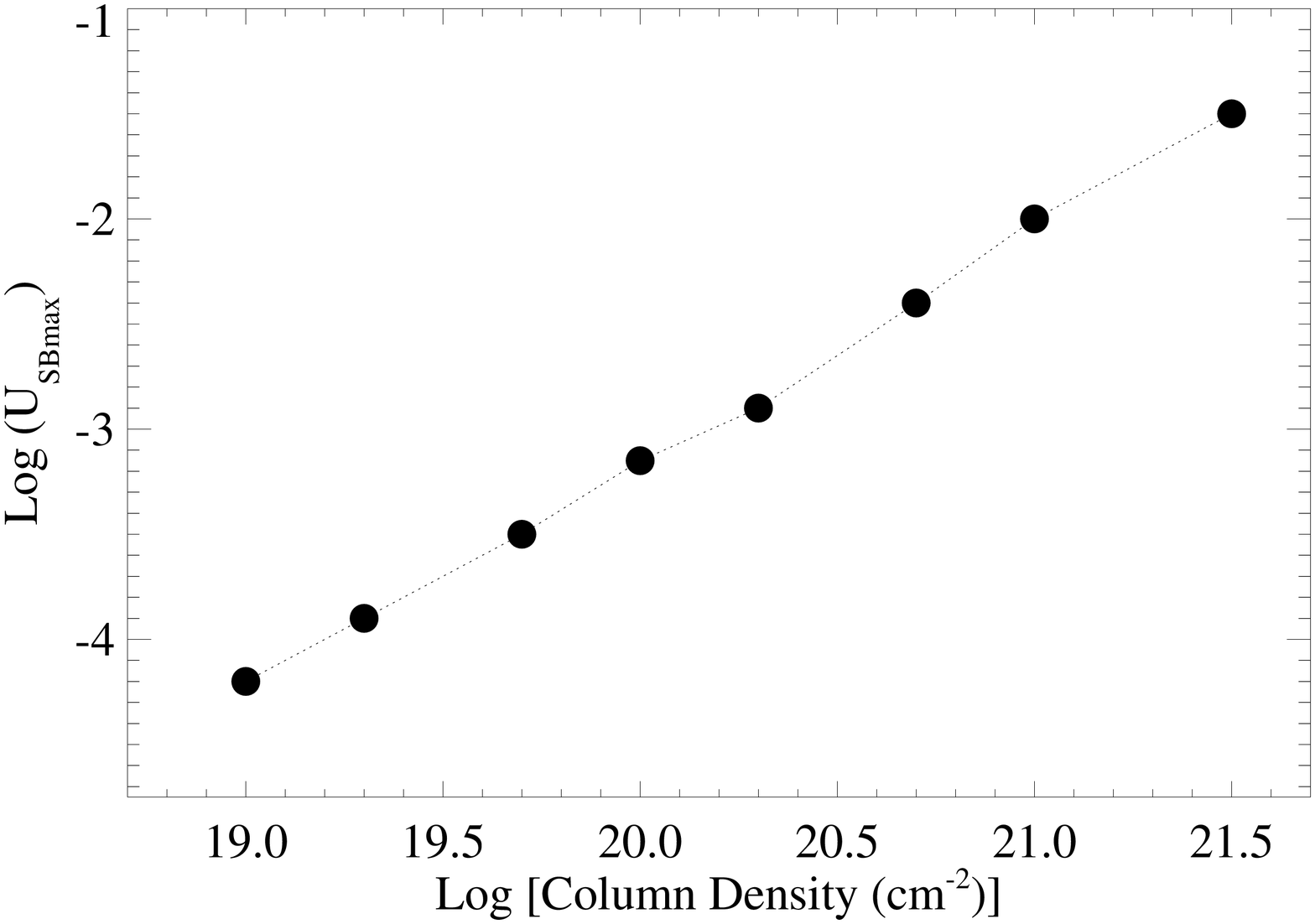} 
\caption{Variation of the peak ionization parameter that a cloud can withstand while remaining primarily neutral, $\rm U_{SBmax}$,as a function of the total column density the cloud. $\rm U_{SBmax}$ is proportional to the column density of the gas over more than 2  orders of magnitude. As the column density of the clouds increase, so does their ability to withstand a stronger radiation field.  \label{cloudy_sb2}}
\end{figure}

 The two tidal-debris dominated groups (HCGs~31 \& 92) show strong evidence for the presence of young stellar populations of 4-10 Myrs old \citep{xu05, torres-flores09, gallagher10}. Here, we explore the possibility of photoionization of the low column density \HI by the escaping ionizing radiation from such young star-forming regions in the vicinity of the \HI in the intragroup medium.  The effect of ionizing radiation on the clouds can be characterized with the ionization parameter, $\rm U_{SB}$, which relates the properties of the starburst (SFR and escape fraction, $f$) and the \HI cloud (distance from the starburst, $d$, and density, $n_{\rm H}$) as \citep[see][]{tumlinson11a} 
 \begin{equation}
 U_{\rm SB}=\frac{10^{-5}}{n_{\rm H}} \rm \Big(\frac{SFR}{10~M_{\odot}~yr^{-1}}\Big)~~\Big(\frac{\it f}{0.01} \Big)~\Big(\frac{20~kpc}{\it d}\Big) ^2 
\end{equation}
The higher the ionization parameter, the greater is the potency of the radiation field to ionize the gas. 
One of the important parameters influencing the ionization parameter is the distance of the cloud from the radiation source. 
The closer the clouds are to the starburst regions, the higher their susceptibility for ionization. 
Therefore, given the properties of the cloud, we can estimate the minimum distance, $\rm D_{HI}$ from the starburst where \HI will be able to withstand ionization and remain primarily neutral i.e. $HI/(HI+HII) > 50\%$. 
Coming back to the case of an expanding tidal structure with initial column density of $\rm 3 \times 10^{21}~cm^{-3}$ and size 1~kpc, $\rm D_{HI}$ can be expressed in terms of the column density as\\
 \begin{math}
  \rm log\Big(\frac{D_{HI}}{kpc}\Big) = -1.198 - 0.5~log~U_{SBmax}~    
\end{math}
 \begin{equation}{\label{eq-U2}}
\rm -0.75~log\Big(\frac{N_H}{3\times 10^{21}~cm^{-2}}\Big) + 0.5~log\Big( \frac{SFR}{10~M_{\odot}yr^{-1}}  \frac{\it f}{0.01}\Big) 
\end{equation}
 where, $\rm U_{SBmax}$ represents the largest value of the ionization parameter that the cloud can withstand while remaining  primarily neutral.
 
In order to explore the relationship between $\rm U_{SB},~ U_{SBmax}$, and the column density of the cloud, we modeled starburst driven photoionization of clouds using the photoionization code, CLOUDY. The modeling assumes the geometry of the clouds to be in the form of a rectangular slab illuminated by a radiation field. A spectrum of a young stellar population of 4~Myrs \citep[from Starburst99;][]{starburst99} was used as the radiation field. At this age the feedback is strongest when the Wolf-Rayet and the O stars dominate the ionizing radiation and the stellar feedback\footnote{However, similar results were obtained for a slightly older population of 10~Myrs.}.
Models were run for clouds with varying total hydrogen column density from $\rm 10^{19}$  to $\rm 10^{21.5}~cm^{-2}$.
Figure~\ref{cloudy_sb} shows the variation of column density of neutral and ionized hydrogen for two of the models. 
The $\rm U_{SBmax}$ is identified by interpolating the column density between consecutive runs where neutral \HI column crosses the 50\% mark. 
Figure~\ref{cloudy_sb2} shows the variation of $\rm U_{SBmax}$ as a function of the column density of the clouds. As expected, as the column density of the clouds increase, so is their ability to withstand stronger radiation fields.

Here we apply the results from our photoionization analysis to determine the minimum distance of the clouds to remain neutral for  two of our groups, HCGs~31 and 92, that are forming stars at a rate of 10~$\rm M_\odot~yr^{-1}$ \citep{gallagher10} and 6.7~$\rm M_\odot~yr^{-1}$ \citep{xu05} respectively. 
Assuming  an escape fraction of ionizing photons as 1\%  \citep{grimes09}, the minimum distance or $\rm D_{HI}$ for clouds of column density  $\rm 3\times 10^{21}~cm^{-2}$ is $\sim$360 and $\sim$290~pc from the star forming regions of HCG~31 and 92 respectively. However, the same for clouds of column density $\rm 2\times 10^{19}~cm^{-2}$ are $\sim$340 and $\sim$280~kpc for HCG~31 and 92 respectively. 

HCG~31 is a very compact group and our data rule out the scenario of the faint gas being more than 300~kpc from the starforming regions. Then how can we understand the presence of faint low column density gas in this system? The likely answer is that faint gas is associated with the HSB (high column density) gas that shields it from the starburst generated radiation field. This scenario is supported by the fact that the diffuse gas found by B10 has the same spectral distribution as the HSB features. On the other hand,  in HCG~92, $\rm D_{HI}$ for a clouds of column density $\rm 5\times 10^{19}~cm^{-2}$ is 77~kpc, which is smaller than the optical size of the group. In fact, our data on HCG~92 show that the diffuse gas in the \HI wing (6300-6500~\kms) is most likely associated with Arc-N, which is 80-100~kpc from the starforming regions. 

We note that besides photoionization, the starburst-driven winds might also lead to heating of the gas via shock ionization. However, the shock ionized gas (if any) will cool back rapidly as cooling time is small \citep{borthakur13} and it is unlikely that the shock ionization would contribute to the long term heating/destruction of the neutral clouds. Detailed hydrodynamical simulations will further help in probing this phenomenon with accuracy. In addition, analytical arguments by B10 indicate that the gas will also survive the conductive heating from the hot IGrM. The contribution of ram pressure stripping and heating  from the hot X-ray intragroup medium likely to be small \citep{rasmussen08}. In fact, even in most X-ray bright groups (except for the most massive ones), \citet{desjardins13} found the X-ray emission is  concentrated mostly around individual galaxies and does not arise from the group medium.

Therefore, it is possible for the diffuse gas to survive for $\approx$500~Myrs or longer.  Again, clouds of higher column densities can have much longer time-scales and can practically survive photionzation from the cosmic UV background. This is supported by the fact that some of the HCGs show large HI structures within which TDG have been detected even a distances of 100~kpc \citep{demello08}. Without detailed hydrodynamical simulations it is not clear the exact survival times for diffuse gas in each of these groups. The conditions such as the temperature of the IGrM, the dynamical mass and effective galaxy densities within the virial radii, total gas content of the IGrM etc. are important parameter in determining the lifetime of the diffuse gas structures. However, we believe that in most groups the basic arguments presented in section 3.5.1 will hold and the diffuse gas will survive for atleast 500~Myrs.

\section{CONCLUSIONS   \label{Sec:conclusion}}

We presented 21cm GBT H I observations of HCGs~31, 44, 68, and 92. The observations were carried out in pointing map modes resulting in a 5x5 map covering an area of 25$^\prime$x25$^\prime$. The pointings were separated by 4$^\prime$. The central pointings were positioned at the center of the groups as defined by \citet{hickson92}.

\begin{itemize}

\item[1.] We detected \HI in all the four groups. The strengths of \HI\ emission and the spectral shapes observed at the central pointing are consistent with our previous single dish measurements. The total \HI masses were found to be $\rm 1.86 \times 10^{10}, 1.82~\times 10^{9}, ~8.32\times 10^{9},~and~2.45\times 10^{10}~M_{\odot} $ for HCGs~31, 44, 68, and 92 respectively.

\item[2.] The tidal debris-dominated groups showed enhanced blending between various spectral features due to the low velocity dispersion among the features. Most of the \HI in these groups exists in single peaked Gaussian-like profile center at the redshifts of member galaxies. HCG~92 shows high velocity wings that connect between two distinct spectral features.
For ISM-dominated groups, most of the \HI lies in disks as confirmed by doubled-horned profiles (44D and 68C). 

\item[3.] We found the \HI observed with the GBT to have the same spatial distribution as the high-surface brightness gas seen with the VLA for HCGs~31 and 68. More than 95\% of the total \HI in these two groups is centrally concentrated as indicated by the minimal difference in flux between the total \HI and the pointing with maximum flux.
The \HI distributions in  HCGs~92 and 44 were spread out and covered most of our maps. The total \HI in these two groups was 23\% and 85\% higher than that observed in the pointing with maximum flux.

\item[4.] We modeled expected the \HI masses for each pointing based on the VLA \HI map. While HCGs~31 and 68 show very little deviation from the modeled values, HCGs~44 and 92 show significant offsets. The faint gas fraction for HCG~44 is large in the Northwest-Southeast direction below the group center. Similarly in HCG~92, we found diffuse gas seen in the \HI ``wing" between velocities 6300-6530~\kms is located in the northwest of the group. The detection of the wing feature in only a few pointings indicates that it is spatially localized and near Arc-N.

\item[5.] The spatial and dynamical similarities between the total (faint+HSB) and the HSB gas indicate that the faint gas detected by the GBT is of tidal origin, like the HSB component. We investigated the effects of the cosmic UV background and the escaping ionizing photons from the star forming regions on the survival of the neutral gas.
We found that the gas will survive and stay primarily neutral for at least 500~Myrs but possibly much longer.

\end{itemize}

\noindent {\bf Acknowledgements}\\

We thank the reviewer for his/her valuable constructive comments.
We thank the staff of the Green Bank Telescope for help during the data acquisition and reduction. In addition, we thank Ching-Wa Yip, Jacqueline van Gorkom, John Hibbard, and Siddharth Srivastava for useful discussions. 
LVM has been supported by Grant AYA2011- 30491-C02-01 and the Junta de Andalu$\acute{c}$ia (Spain) P08-FQM-4205 and TIC-114.

\bibliographystyle{apj}	        
\bibliography{myref_bibtex}		


\begin{deluxetable} {lccc ccc  l}

\tablenum{1}
\tabletypesize{\scriptsize}
\tablecaption{Properties of our Sample. \label{tbl-sample}}
\tablewidth{0pt}
\hspace{-1cm}
\tablehead{
\colhead{Group} & \colhead{Redshift} &\colhead{Velocity} & \colhead{Distance}& \colhead{Group diameter}  & \colhead{Map center} & \colhead{Scale}   & \colhead{Previous HI imaging studies} \\
\colhead{}          & \colhead{z$_{gp}$} &\colhead{$V_{gp}$} & \colhead{(L$_{lum}$)}& \colhead{$\Theta_{gp}$ ~~~D$_{gp}$}  & \colhead{~~~R.A.~~~~~~Dec~~} & \colhead{}   & \colhead{} \\
\colhead{} &\colhead{} & \colhead{(km~s$^{-1}$)} & \colhead{(Mpc)} & \colhead{(arcmin) ~~(kpc)} & \colhead{} & \colhead{(kpc/$^\prime$)}  & }
\startdata
HCG~31 & 0.01347 & 4039 & 58.3 &   4.2$^a~~$~~  69  &  05:01:38.43~-04:15:25.0   & 16.5  & \citet{VM05} \\
HCG~44 & 0.00460 & 1379 & 19.8 & 16.4~~    ~  93  &  10:18:00.45~+21:48:42.9  &  5.7  & \citet{williams91, serra13} \\
HCG~68 & 0.00800 & 2310 & 34.5 &   9.2~~    ~~  91  &  13:53:40.99~+40:19:09.5  & 9.8  & \citet{tang08}  \\
HCG~92$^b~~$~~  & 0.02150 & 6446 & 93.6 &   3.2~~     ~~ 84  &  22:35:57.52~+33:57:37.4   & 26.1 &\citet{williams02} 
\enddata
\tablenotetext{a}{The group diameter for HCG~31 was revised from the original value of $\theta_{gp}=\rm0.9^{\prime}$  \citep{hickson82} to $\theta_{gp}=\rm4.2^{\prime}$ by \citet{VM05} based on VLA HI observations.}
\tablenotetext{b}{HCG~92 is known as Stephan's Quintet. Our observations do not include the foreground galaxy HCG~92A (aslo known as NGC 7320) at a redshift of 0.002622 ($v=\rm786~$\kms).}
\end{deluxetable}

\begin{deluxetable} {lccc cccc  l}
\tablenum{2}
\tabletypesize{\scriptsize}
\tablecaption{Details of the Observations. \label{tbl-observations}}
\tablewidth{0pt}
\tablehead{
\colhead{Group} & \colhead{$t_{integration}^a$} &\colhead{$\Delta~V_{HI}^b$} & \colhead{HI mass limit$^c$} \\
\colhead{} &\colhead{(s)} & \colhead{(km~s$^{-1}$)} & \colhead{(Log[M$_\odot$])}  }
\startdata

HCG~31 & 480 & 3900-4250  & 8.69-8.86 \\
HCG~44 & 260 & 1000-1750   & 7.88-7.91\\
HCG~68 & 420 & 2100-2500    & 8.30-8.46 \\
HCG~92 & 300 & 5950-6950 & 9.26-9.33 
\enddata
\tablenotetext{a}{On-source integration per pointing.}
\tablenotetext{b}{The range of velocities over which the HI emission was detected. There may be multiple components found within this velocity window.}
\tablenotetext{c}{The values show the range of M$_{5\sigma}$ for all the pointings around each of the HCG. The HI mass was estimated assuming a line width of $\Delta V_{HI}=$400~\kms in line-free spectral windows. }
\end{deluxetable}

\begin{deluxetable}{| l |c |c |c |c|c|}
\tablenum{3}
\tabletypesize{\scriptsize}
\tablecaption{Angular distances (in arcminutes) of pointings from the center of the map. \label{tbl-rho_grid}}
\tablewidth{0pt}
\tablehead{
\colhead{} &  \colhead{1} & \colhead{2} & \colhead{3} & \colhead{4} & \colhead{5} }
\startdata
    A  & 11.3 & 8.9 & 8.0 & 8.9 & 11.3 \\ \hline
    B  &   8.9  & 5.7 & 4.0 & 5.7 & 8.9 \\ \hline
    C  &  8.0  & 4.0 & 0.0 & 4.0 & 8.0 \\ \hline
    D  &   8.9  & 5.7 & 4.0 & 5.7 & 8.9 \\ \hline
    E  &   11.3 & 8.9 & 8.0 & 8.9 & 11.3 
\enddata
\end{deluxetable}

\begin{deluxetable}{| c |r |r |r |r |r|}
\tablenum{4}
\tabletypesize{\scriptsize}
\tablecaption{HI masses for each pointing.  \label{tbl-hi_mass} }
\tablewidth{0pt}
\tablehead{
\colhead{Position} & \colhead{HCG~31} & \colhead{HCG~44} & \colhead{HCG~68} & \colhead{HCG~92} \\
\colhead{}         & \colhead{(Log[M$_\odot$])}  & \colhead{(Log[M$_\odot$])}  & \colhead{(Log[M$_\odot$])}  & \colhead{(Log[M$_\odot$])}  }
\startdata
      A1   &  $<$8.56  & -  & $<$8.30  & $<$9.27\\
      A2   &  9.06  & -  & 8.73  & 9.65\\
      A3   &  9.29  & 8.58  & 9.37  & 9.66\\
      A4   &  9.09  & 8.64  & 9.55  & $<$9.29\\
      A5   &  $<$8.63  & 8.26  & 9.25  & $<$9.28\\
      B1   &  8.88  & 8.25  & $<$8.35  & 9.84\\
      B2   &  9.76  & 8.52  & 9.19  & 10.17\\
      B3   &  10.01  & 8.87  & 9.75  & 10.18\\
      B4   &  9.80  & 8.99  & 9.91  & 9.60\\
      B5   &  9.02  & 8.72  & 9.64  & $<$9.27\\
      C1   &  9.37  & 8.35  & $<$8.35  & 10.13\\
      C2   &  10.01  & 8.79  & 9.09  & 10.30\\
      C3   &  10.26  & 8.97  & 9.62  & 10.20\\
      C4   &  10.04  & 8.98  & 9.80  & 9.59\\
      C5   &  9.24  & 8.72  & 9.58  & $<$9.26\\
      D1   &  9.25  & $<$7.91  & $<$8.39  & 9.64\\
      D2   &  9.90  & 8.46  & $<$8.40  & 9.68\\
      D3   &  10.11  & 8.70  & 9.08  & 9.77\\
      D4   &  9.83  & 8.87  & 9.40  & $<$9.27\\
      D5   &  8.90  & 8.50  & 9.13  & $<$9.27\\
      E1   &  9.10  & $<$7.89  & $<$8.40  & $<$9.28\\
      E2   &  9.42  & $<$7.88  & $<$8.41  & $<$9.28\\
      E3   &  9.47  & 8.25  & $<$8.45  & $<$9.29\\
      E4   &  9.04  & 8.63  & $<$8.45  & $<$9.32\\
      E5   &  $<$8.84  & 8.59  & 8.44  & $<$9.32\\
      \hline
  &  &   &    & \\
      Total$^{a}$ &  10.27    & 9.26     &  9.92      &10.39
      \enddata
\tablenotetext{a}{Estimated from composite HI spectra shown in Figure~3.}
\end{deluxetable}

\end{document}